\def\etal{et al.~}
\def\msun{{\rm M}_\odot}
\def\H0{$H_0$ = 100 {\it h} km s$^{-1}$ Mpc$^{-1}$}

\documentclass[preprint]{aastex}

\usepackage{psfig}
\usepackage{graphics}
\usepackage{natbib}

\slugcomment{Accepted for publication in the Astronomical Journal}

\shortauthors{Galaz et al.}
\shortauthors{LSBGs in the near-IR}

\begin{document}

\title{Properties of Low Surface Brightness Galaxies and Normal Spirals in the Near-IR}

\author{Gaspar Galaz\altaffilmark{1}}
\affil{Departamento de Astronom\'{\i}a y Astrof\'{\i}sica, 
Pontificia Universidad Cat\'olica de Chile, Casilla 306, 
Santiago 22, Santiago, Chile}
\email{ggalaz@astro.puc.cl}

\author{Julianne J.\ Dalcanton\altaffilmark{2}}
\affil{Department of Astronomy, Box 351580, University of 
Washington, Seattle WA 98195, USA} 
\email{jd@astro.washington.edu}

\author{Leopoldo Infante}
\affil{Departamento de Astronom\'{\i}a y Astrof\'{\i}sica, 
Pontificia Universidad Cat\'olica de Chile, Casilla 306, 
Santiago 22, Santiago, Chile}
\email{linfante@astro.puc.cl}

\author{Ezequiel Treister\altaffilmark{1}}
\affil{Departamento de Astronom\'{\i}a, Universidad de Chile, Casilla 36-D, 
Santiago, Chile \\ and \\ Department of Astronomy, 
Yale University, P.O. Box 208101 New Haven, CT 06520-8101 USA}
\email{treister@astro.yale.edu}

\altaffiltext{1}{Visiting Astronomer, Las Campanas Observatory.}
\altaffiltext{2}{Alfred P. Sloan Research Fellow}

\begin{abstract}
  We present results for $J$ and $K_s$ near-IR imaging data on a large
  sample of 88 galaxies drawn from the catalogue of Impey et
  al. (1996).  The galaxies span a wide range in optical and IR
  surface brightness and morphology (although they were drawn from a
  catalog constructed to identify low surface brightness galaxies,
  hereafter LSBGs).  They were also selected to include very low and
  high HI mass galaxies in order to ensure that they span a wide range of
  evolutionary states.

  The near-IR data unveils many features of LSBGs not seen before in
  the optical.  First, a high fraction of the observed LSBGs are very
  luminous in the near-IR, indicating that they have a well developed
  old stellar population, and that older LSBGs are more frequent in
  the universe than data from optical bands suggested. 
  Second, the near-IR morphologies are
  often quite different than seen in the optical.  Many diffuse LSBGs
  that are apparently bulgeless when observed in blue bands, instead
  exhibit nuclei in $J$ and $K_s$ bands.  Third, we find significant
  trends between the near-IR morphologies of the galaxies and their
  ratio of HI mass to near-IR luminosity.  Fourth, we find no trend in
  disk surface brightness with absolute magnitude, but significant
  correlations when the bulge surface brightness is used.  Finally, we
  find that the formation of a bulge requires a galaxy to have a total
  baryonic mass above $\sim\!10^{10}\msun$.

  A wide variety of other correlations are explored for the sample.
  We consider correlations among morphologies, surface brightnesses,
  near-IR colors, absolute magnitudes, and HI masses.  In addition,
  using previous results by \citet{bell2001}, we convert the galaxies'
  near-IR luminosities to stellar masses, using color-dependent
  stellar mass-to-light ratios.  This allows us to consider
  correlations among more fundamental physical quantities such as the
  HI mass, the stellar mass, the total baryonic mass, the gas mass
  fraction, the mass surface density, and the metallicity (via the
  highly metal sensitive color index $J-K_s$).

  We find that the strongest of our correlations are with the ratio of
  HI mass to total baryonic mass, M$_{HI}$/M$_{baryonic}$, which
  tracks the evolutionary state of the galaxies as they convert gas
  into stars, and which ranges from 0.05 up to nearly 1 for the
  galaxies in our sample.  We find strong systematic trends in how the
  metallicity sensitive $J-K_s$ color becomes redder with decreasing
  M$_{HI}$/M$_{baryonic}$, as would be expected for ``closed box''
  models of chemical enrichment.  However, the increased scatter with
  increasing gas mass fraction and decreasing galaxy mass suggests
  that gas infall is increasingly significant in the gas rich lower
  mass systems.  We argue that the overall range in $J-K_s$ color
  argues for at least a factor of 20 change in the mean stellar
  metallicity across the mass range spanned by our sample.  We also
  see strong trends between M$_{HI}$/M$_{baryonic}$ and central
  surface density, suggesting that increased star formation efficiency
  with increasing gas surface density strongly drives the conversion
  of gas into stars.

\end{abstract}


\section{Introduction}

Low surface brightness galaxies (LSBGs) have been a subject of
increasing interest in the last two decades.  Although an initial
study by Freeman (1970) suggested that the central surface
brightnesses of disk galaxies in the Hubble sequence (Sa-Sb-Sc) fell
within a rather narrow range, \citet{disney76} pointed out that the
lack of low surface brightness galaxies may in fact be a selection
effect due to the difficulty in discovering galaxies of very low
surface brightness, as had previously been recognized by
\citet{zwicky57}.  Indeed, many surveys have since turned up large
numbers of LSB galaxies.  In practice, ``low surface brightness'' has
come to mean galaxies whose central surface brightness is fainter than
22.0 mag arcsec$^{-2}$ in the $B$ band-pass (i.e. more than $1\sigma$
outside of the narrow range defined by \citet{freeman1970} of
$\mu_0 = 21.7 \pm 0.3 B$-mag/arcsec$^{2}$).  However, while many galaxies
have been found below the ``Freeman value'' of surface brightness,
disk galaxies with surface brightnesses that are significantly higher
than those of normal spirals do not appear to exist. This is shown
most clearly in \citet{courteau96}, who find that there is a rather
well-defined upper cut-off at a central surface brightness in $R$ of
$\sim 20.08$ mag arcsec$^{-2}$. 

It was originally thought that all galaxies with a low surface
brightness were early or late-type dwarfs \citep{vandenbergh59}.
However, radial velocity observations by \citet{fisher75} showed that
some galaxies with low surface brightness are actually quite large and
luminous \citep{mcgaugh92, bothun97}, for example LSBGs like Malin 1
type.  Indeed, most surveys have revealed LSBGs with disk scale lengths
comparable to the Milky Way (although these angular diameter limited surveys
are strongly biased toward finding the physically largest galaxies).

Based upon surveys to date, LSBGs do seem to have different properties
than those of brighter galaxies (possibly indicating different
evolutionary histories).  The observed trends suggest that LSBGs are
in large part a continuous extension of the population of normal disk
galaxies, reaching to higher angular momenta and lower masses
\citep{dalcanton97}.  Previous studies \citep{longmore82} showed that
LSBGs are much more gas rich (in terms of $M_{gas}/M_{stars}$) and
bluer than ``normal'' late type galaxies \citep{mcgaugh92, deblok95,
  sprayberry95}.  When combined with their observed low gaseous
metallicity \citep{mcgaugh94, deblok97}, the evidence suggest that
LSBGs have had a historically low star formation rate and are
relatively unevolved systems compared to the bulk of normal spirals.
HI observations \citep{vanderhulst93, deblok96} show that LSBGs have
extended disks with low gas surface densities and high M/L ratios,
further confirming the unevolved nature of LSBGs.  Even external
influences seem to have had little effect in speeding up the evolution
of LSBGs \citep{mcgaugh97}.

Despite the tremendous progress in this field, there are still
significant uncertainties in many of the above results.  First, while
the data suggests a young stellar age for LSBGs overall, the spread in
the integrated optical colors of LSBGs is large, suggesting that they
may actually have traveled along diverse evolutionary paths.  In
particular, because of the low surface brightness of the underlying
population, only a small fraction of the total number of stars is
needed to make the colors significantly blue, and thus a small change
in the star formation rate can drastically change the apparent mean
age of the stellar population as inferred from {\emph{optical}}
observations.  Second, while emission line measurements have
revealed the metallicity of the current gas phase, they have not
measured the metallicity of the underlying stellar population. Given
the large gas reservoirs observed in LSBGs and the possibility of
substantial late time gas infall, it is not clear that the current gas
phase metallicity is closely related to the actual metallicity of the
stars.  Third, the total stellar mass in these galaxies (and their
contribution to dynamics) is rather uncertain, given the large
uncertainties in the stellar mass-to-light ratio at optical
wavelengths.  Finally, recent modeling of LSBGs predict that from the
assumption that blue LSBGs are currently undergoing a period of
enhanced star formation, there should exist a population of {\em red},
non-bursting, quiescent LSBGs \citep{gerritsen98, beijer99, bell2000, 
vandenhoek2000}.
These galaxies should then also be metal-poor and gas-rich, and share
many of the properties of the LSBGs observed in the optical.

In all of these areas, we can greatly improve the observational
constraints by undertaking a systematic study of the properties of
LSBGs and normal spirals in the near-IR.  Combined near-IR and
(eventually) optical observations are a perfect probe to help to
clarify the true stellar masses of LSBGs, to disentangle the roles
played by the age and metallicity, and to understand their place in
the continuum with normal high surface brightness (HSB) spirals.

We expect that contrary to initial impressions, LSBGs should be readily
detectable in the near-IR.  Although much of the data suggests that
the mean stellar population of LSBGs are probably young, we know from
the very red integrated colors of some LSBGs (such as the giant Malin-type
objects) that many LSBGs contain a significant {\em old population of
  stars}, and were therefore not formed recently.  This old population
should dominate the near-IR (which is a good tracer of red giant and
lower mass stars) and thus should be detectable with NIR observations.
We also should be able to easily identify LSBGs with bulges, as these
are among the reddest members of the LSBG population \citet{mcgaugh92}.
There are also indications that the most massive LSBGs, such as the
super-giant galaxies Malin 1 and Malin 2, are redder than the majority
of LSBGs, with luminosities that are comparable to those of normal
spirals.  Thus, near-IR observations should be particularly sensitive
to the most massive end of the LSBG population.

While valuable, previous studies of the IR properties of galaxies in
the low surface brightness regime have been limited due to the
difficulty of the observations.  \citet{knezek94} observed a
preliminary sample of LSBGs in $J$, $H$ and $K$, but the sample was
very small and biased toward massive galaxies.  \citet{bergvall99}
have observed 14 blue LSBGs in $J$, $H$ and $K$ and combined with
optical photometry, arriving to the result that many properties of
LSBGs observed in the optical are reproduced in the near-IR. In
particular, they observe the same morphologies in $B$ and $J$,
measuring weak optical/near-IR color gradients, a fact which they
interpret as a low dust content.  

Near-IR observations have also been used to probe the stellar
population differences between LSBGs and normal spirals.
\citet{bell2000} undertook a combined optical and IR study of 26
LSBGs, including 5 from his earlier thesis work in \citet{bell99}, and
examples of both normal blue LSBGs and the rarer class of red LSBGs.
They compare LSBG average ages and metallicities with their physical
parameters and find that blue LSBGs are well described by models with
low, roughly constant star formation rates, whereas red LSBGs are
better described by a ``faded disc'' scenario.  Among the larger
population, strong correlations are seen between an LSBG's star
formation history and its $K$-band surface brightness, $K$-band
absolute magnitude and gas fraction.  These correlations are
consistent with a scenario in which the star formation history of an
LSBG primarily correlates with its surface density and its metallicity
correlates with both its mass and its surface density.  Another larger
study involving IR data of spiral galaxies was made by \citet{dejong96},
but included only a very small number of true LSBGs.

The lack of much previous work, and the availability of relatively
large IR array detectors, both open the possibility to systematically
observe a large sample of LSBGs and normal spirals in the near-IR, to
derive fundamental parameters related to the origin of these galaxies
and to their relation to normal galaxies.  Based upon the large
expected rewards and the limited existing data, a systematic near-IR
study of LSBGs is needed in order to complement existing optical data.
We undertake such a study in this paper, which is organized as
follows. In \S\ref{samplesec} we present the sample used in this work.
\S\ref{obssec} briefly describes our observations, data reductions,
photometric calibrations, galaxy photometry, and surface photometry.
In \S\ref{resultsec} we analyze and discuss the most important
results, including galaxy photometry, central surface brightnesses and
scale lengths, the relationship between HI-mass and near-IR
luminosities, and the relationship between stellar mass-luminosity
ratios and baryonic mass.  Finally, we conclude in
\S\ref{conclusionsec}.

\section{Sample}                \label{samplesec}

All galaxies were initially selected from the catalogue of large
angular size low surface brightness galaxies compiled by
\citet{impey96} from the APM plates, consisting of 516 LSBGs with
effective radius larger than $\sim 10$ arcsec.  Some constraints were
applied to optimize the galaxies' observability and their intrinsic
stellar and gas content.  First, a surface brightness cut was applied
to ensure detectability at the 40-inch Swope telescope and at the
100-inch du Pont telescope, both located at Las Campanas Observatory
(LCO). The corresponding central surface brightness cut-off was set to
$\mu_0(B) = 22.0$ mag arcsec$^{-2}$ for the galaxies observed with the
40-inch telescope, and to $\mu_0(B) = 23.5$ mag arcsec$^{-2}$ for the
galaxies observed with the 100-inch telescope. These limits are set
considering (1) the corresponding limiting $K_s$ magnitudes for the
40-inch and the 100-inch telescopes both equipped with a NICMOS3
HgCdTe array: 19.0 mag and 20.5 mag, respectively, and (2) the
observed average color $<B - K> \sim 3.0$ for LSBGs \citep{tully97}.
We applied additional selection criteria in declination, $\delta \le
5^\circ$, in order to observe galaxies from LCO.  We then randomly
sampled from the resulting catalog to generate a sample which
would be observable during the one year of allotted telescope time.
Finally, we restricted the sample to two ranges in HI mass, selecting
galaxies with either log$[M_{HI}/M_{\sun}] \le 9.0$ (sample A) or
log$[M_{HI}/M_{\sun}] \ge 9.5$ (sample B).  This last selection
criterion allows us to capture LSBGs in the maximum range of
evolutionary states (i.e. high and low gas-to-star ratios) and total
masses.

There are 112 galaxies from the \citet{impey96} catalogue satisfying 
the above constraints. Of these, 65 galaxies were observed with the 
40-inch telescope and 47 galaxies with the 100-inch telescope. 

In order to have an optimal wavelength baseline, we observed all
galaxies through both the $J$ and $K_s$ near-infrared filters (the 
latter defined by \citet{persson98}).  Due to the critical
importance of accurate surface brightness measurements, and given the
low surface brightness of these galaxies, we chose not to use the $H$
band.  The strong and variable OH emission which dominates the sky
background at $H$ is not easy to remove, and could significantly
affect the apparent surface brightness of the galaxies. In addition,
prominent fringing appears in the $H$ band, putting additional hurdles
to deriving the correct surface brightness flux.  The $J$ band, on the
other hand, is stable during time scales of $\sim 20$ minutes, and the
$K_s$ bands is stable enough typically for time scales shorter than 5
minutes, when the outside temperature is below 15$^\circ$C
\citep{galaz2000}.  Unlike many earlier near-IR studies of nearby galaxies,
we have sufficient sky background in our images to do a reasonable job
of sky subtraction.  Most galaxies are much smaller than the $\sim 2
\times 2$ arcmin FOV of both telescopes, with just a few having extents
larger than 1.5 arcmin in the $B$ band.

\section{Observations and Reductions}           \label{obssec}

All galaxies were observed by GG and EZ during photometric nights at
LCO, using the 40-inch Swope telescope and the 100-inch du Pont
telescope. Seven runs of 3 to 7 nights were allocated for this
project. Observations started on February 23rd 1999 and finished on
October 23rd 1999.  A NICMOS3 HgCdTe array was used in both
telescopes, but with different pixel scales: 0.60 arcsec/pix at the
40-inch, and 0.348 arcsec/pix at the 100-inch. The corresponding FOV
were $2.56^\prime \times 2.56^\prime$ and $1.48^\prime \times
1.48^\prime$.  Observation procedures were the same as used in
\citet{galaz2000}, and consist of a sequence of short
dithered exposures for all the scientific targets, including the
standard stars. Due to the low surface brightness which characterizes
these galaxies, exposure times were calculated based on the known blue
surface brightness, extrapolated to the $J$ and $K_s$ band, and
corrected for the additional contribution of the thermal noise and sky
emission.  Typical total exposure times were between 20 min and 45 min
in $J$ and between 40 min and 75 min in $K_s$.

Considering the low surface brightness of galaxies, special care was
taken with flat-fielding, in order to avoid undesirable artificial
spatial fluctuations.  A large series of dome-flats were taken for
each run, as well as sky-flats every sunset. Additionally, the raw
images of each night were used to construct super-flats, by combining
dithered images, using a median sky value obtained after a sigma
clipping algorithm to discard cosmic rays, bad pixels, and any
residual contribution from the galaxies themselves. Usually, no fewer
than 80 images per filter were used to construct nightly super-flats,
and only images where the galaxies were small compared the chip size
were included in the image stack.  Afterward, dome-flats, sky-flats
and super-flats were compared. The largest differences between sky and
dome-flats were $\sim 1.5\%$.

For image reduction, we used a modified version of
DIMSUM\footnote{DIMSUM is the Deep Infrared Mosaicing Software package
  developed by Peter Eisenhardt, Mark Dickinson, Adam Stanford, and
  John Ward.}. The reduction procedure is completely analogous to that
described in \citet{galaz2000}, to which the reader can refer
for details. Out of the 112 original selected LSBGs, we
successfully observed 88; 39 from sample A (low HI mass) and 49
from sample B (high HI mass).  

\subsection{Calibrations}

In order to derive color information we calibrated our magnitudes
system using the standard star system defined by \citet{persson98}.
For the standard star photometry IRAF's\footnote{IRAF is distributed
  by the National Optical Astronomy Observatories, which are operated
  by the Association of Universities for Research in Astronomy, Inc.,
  under cooperative agreement with the National Science Foundation.}
``DAOPHOT'' software was used. We adopted a 10 arcsec aperture radius for
all the standard stars, which corresponds to the same aperture used by
\citet{persson98}.  Typical seeing was 1.0 arcsec. No less than 6
standards were measured each night, which allow us to derive
calibrated $J$ and $K_s$ magnitudes.  The transformation equations
were
\begin{eqnarray}
J   & = & j + a_1 - X \xi_j + a_2(j - k_s)   \\
K_s & = & k_s + b_1 - X \xi_{k_s} + b_2(j - k_s),
\label{transformations}
\end{eqnarray}
where capitals denote standard magnitudes, small characters denote
instrumental magnitudes, $a_1$ and $b_1$ are the zero points, $a_2$
and $b_2$ are the color terms, and $\xi_j$ and $\xi_{k_s}$ are the
extinction coefficients in $J$ and in $K_s$ respectively. $X$ is the
mean airmass of the observation. 

We have computed the transformation parameters of equations (1) and
(2) using ``photcal'', included in IRAF.  When measuring the above
coefficients, color terms and zero points were initially computed
independently for each night, verifying that they changed by only a
small amount (less than 5\%) from night to night; because these terms
depend only upon the telescope/mirror-instrument-filter combination,
they should be constant over each run.  We then adopted a single color
term and zero point for the entire run (based upon the night-to-night
average), and re-computed the extinction coefficients for each night
separately.  The values obtained were very close to the expected
values $\xi_j = 0.10$ and $\xi_{k_s} = 0.08$ in \citep{persson98}.
The resulting average values typical of all our runs are shown in
Table \ref{calib}.  The typical color term is relatively small, and
given the observed values of $J - K_s$ (between 0.10 and 0.80), the
contribution of this term is $\sim 0.15$ in $J$, and therefore may be
marginally important when computing the spatial distribution of colors
through a given galaxy.

\placetable{calib}

\subsection{Galaxy Photometry}  

Given the heterogeneous luminosity profiles of the LSBGs, several
different techniques were initially explored to optimize measurement
of the apparent magnitudes: (1) circular apparent magnitudes, using
DAOPHOT and the growth curve method, (2) total magnitudes within
elliptical apertures, computed using SExtractor \citep{bertin96}.
SExtractor also computes isophotal areas which allow one to compute
mean surface brightness within a specified isophote level. A
comparison between these two techniques allows us to conclude that a
more robust measurement of the flux is obtained with SExtractor
magnitudes, given the elongated shape of many of the LSBGs in the
sample (see Figures \ref{mosaics1} and \ref{mosaics2}).  In LSBGs with
small ellipticities, DAOPHOT and SExtractor magnitudes differ by $\sim
8\%$. In what follows, we use only magnitudes computed with SExtractor
on elliptical apertures.  Magnitudes were determined within an
isophotal level set by the quality of the data, at a 3$\sigma$ level
above the sky; the resulting sizes of the elliptical apertures are listed in
Table~\ref{phot1}. The major uncertainty arises from the sky level,
with a mode of $\sim 5$\% in $J$ and $\sim 8$\% in $K_s$.  We have
attempted to compare our derived magnitudes with those from the 2MASS
survey, but found no overlap in the galaxies studied.

\placetable{phot1}

Table \ref{phot1} shows the $J$ and $K_s$ magnitude and color, for the
observed LSBGs. The heliocentric radial velocity from the 21-cm line,
the size of the elliptical aperture used for photometry, and the HI
mass content from \citet{impey96} are also shown.  
The absolute magnitudes $M$ in Table \ref{phot1} were computed as
\begin{equation}
M = m - 5\log\left[\frac{(cz+v_{pec})}{H_0}\right] - 25 
\label{abs_magnitude}
\end{equation}
where $m$ is the apparent magnitude, $H_0$ is the Hubble constant
expressed in km sec$^{-1}$ Mpc$^{-1}$ (we use $H_0 = 75$ km sec$^{-1}$
Mpc$^{-1}$ in this paper), and $v_{pec}$ is the Virgocentric peculiar
velocity in km/s. Throughout this paper we use $v_{pec} = 600$ km/s,
as quoted by \citet{kraan1986,giovanelli1998}. This corrective term is
important for galaxies with $cz \la 2,000$ km/s.  It is worth noting
that this correction {\em does not} depend on passband, and therefore
does not affect colors.  Equation \ref{abs_magnitude} is a good
approximation up to a recessional velocity $cz \sim 25,000$ km
sec$^{-1}$, where cosmological corrections become important and the
parameter $q_0$ must be taken into account (from Table \ref{phot1} we
see that our largest velocity is $\sim 23,000$ km sec$^{-1}$).  In
equation \ref{abs_magnitude} we have neglected $k$-corrections. At the
LSBGs redshifts $k$-corrections are smaller than our photometric errors,
and therefore are negligible.  Note also that heliocentric radial
velocity is measured from the HI emission (Table \ref{phot1}),
slightly different from the optical heliocentric radial velocity.
However, differences are smaller than 100 km/sec, and therefore
changes in absolute magnitudes are no larger than 0.05 mag.

As can be seen from Table \ref{phot1}, LSBGs do not
necessarily have faint absolute magnitudes in the near-IR.
Instead, some of them appear to be as bright in the near-IR as normal
high surface brightness galaxies.  This immediately suggests that
they do have a significant population of old stars. 

\subsection{Measuring Central Surface Brightnesses and Scale Lengths}

Measurements of surface brightness and scale lengths give critical
information on the light density and physical size of galaxies.  We
now use surface brightness fitting to measure the central surface
brightness (as an indicator of the central light concentration) and
scale length of both the disk and bulge components of the galaxies in
our sample.  The reader must recall that the surface brightness
difference we measure for the same extended source located at
different redshifts $z_1 < z_2$ is given by
\begin{equation}
\Delta\mu = 10 \log\left(\frac{1 + z_2}{1 + z_1}\right),
\label{delta_mu}
\end{equation}
where $\Delta\mu$ is expressed in mag arcsec$^{-2}$. In our case, our
smallest radial velocity is $\sim$850 km sec$^{-1}$ ($z \sim 0.003$),
and our farthest galaxy is located at $\sim 29,000$ km sec$^{-1}$ ($z
\sim 0.1$). This implies that the surface brightness difference for an
object with a given surface brightness at $z_1$, but measured at
$z_2$ is, at most, $\Delta\mu = 0.4$ mag arcsec$^{-2}$. However, at
$\pm3\sigma$ level around the average redshift, we have $v_1 \sim
5,000$ km sec$^{-1}$ and $v_2 \sim 13,000$ km sec$^{-1}$, which yields
$<\Delta\mu> \sim 0.08$ mag arcsec$^{-2}$, which is inside the
magnitude error bars. Therefore our surface magnitude estimates are
not strongly modified by the range of distance included in the sample.

We adopt the same approach as given by \citet{beijer99}, and fit our
data to the following parametric profile
\begin{equation}
I(r) = I_0 \exp (-r/h),
\label{profile_int}
\end{equation}
or in terms of magnitudes per square arc-second
\begin{equation}
\mu(r) = \mu_0 + 1.086 \times (r/h).
\label{profile_mags}
\end{equation}
In equations \ref{profile_int} and \ref{profile_mags} $I_0$ and
$\mu_0$ are the central surface brightness, the latter measured in mag
arcsec$^{-2}$.  The parameter $h$ is the scale length, measured in
arcsec and transformed to physical units (kpc). Similar to
\citet{beijer99}, we initially fit bulges using de Vaucouleurs
profiles \citep{devaucouleurs48}. However, the profile was not a good
match to the bulge light profile, and poor fitting results were
obtained.  Following \citet{beijer99}, we then switched to fitting the
bulge with an exponential light profile as well, yielding much better
fits.  An example bulge and disk fit is presented in Figure
\ref{lsb059_profiles}.  The double exponential profile provides an
adequate fit to the data over a large range in radius, though slight
deviations suggest that an intermediate S\'ersic profile for the bulge
might provide a slightly improved fit.  We find that most of the
galaxies require a bulge component in the near-IR (i.e.\ they show
substantial curvature in the inner parts of a $\mu$ vs $r$ plot), but
some do not and represent completely disk dominated galaxies. 

\placefigure{lsb059_profiles}

\section{Results and Discussion}        \label{resultsec}

\subsection{Morphologies and Surface Brightnesses} 

A mosaic with reduced and combined
$J$ images is shown in Figures \ref{mosaics1} (sample A, low HI mass
LSBGs) and \ref{mosaics2} (sample B, high HI mass LSBGs).  Typical image
sizes are $\sim 1.5^\prime \times 1.5^\prime$ arcmin.  
Both Figures \ref{mosaics1} and \ref{mosaics2} and our surface
brightness decompositions reveal a number of significant results for
the morphologies and surface brightnesses of our sample.

\placefigure{mosaics1}
\placefigure{mosaics2}  

First, many of the LSBGs classified as late types based upon the
optical $B$ band image instead show a dramatic bulge in the near-IR
(confirming the presence of prominent near-IR bulges noted by
\citet{beijer99}).  This result differs from that obtained by
\citet{bergvall99} and \citet{bell2000}, who found similar
morphologies in the $B$ and $J$ bands, but for a much smaller sample
of galaxies.  Examples of galaxies with different optical and near-IR
morphologies (such as in the presence of a bulge in the near-IR) are
the cases of LSBG \#100 and LSBG \#336 (for example, their $J$ images
in Figure \ref{mosaics1}). The fact that some LSBGs have high surface
brightness bulges ($\mu_0(J) < 17.5$ mag arcsec$^{-2}$) has been
commented by \citet{beijer99}, who noted that many LSBGs could be
regarded as high surface brightness bulges embedded in low surface
brightness disks.  Our results support such a hypothesis.  In other
words, while some galaxies have been classified as LSBGs on the basis
of their {\emph{disk}} properties, they are not low surface brightness
overall.  This difference is particularly striking in the near-IR,
where the surface brightness more closely reflects the stellar surface
density.  Moreover, as discussed by \citet{impey96}, many of the
galaxies in the catalog from which our sample are drawn are not true
LSBGs at all, and are instead indistinguishable from normal spirals in
the optical.  We find a similar result for our near-IR data.

Second, in Figure \ref{MJ_mu0} we plot the infrared luminosity of the
galaxies vs the central surface brightness of the disk component.  The
figure shows that there is no correlation between these quantities and
that almost all of the absolute magnitude vs surface brightness
parameter space is covered. The upper left part of the distribution
(large, luminous but LSB galaxies) is as populated as the the lower
right part of the distribution (small, quite compact galaxies). In our
sample there is no bias against large and faint, nor small but
luminous galaxies.  We also observe in Figure \ref{MJ_mu0} a sharp
cutoff towards bright surface brightnesses (at $\mu_J(disk)\sim 17.5$
mag arcsec$^{-2}$).  This cutoff represents a real physical cutoff
(i.e.\ nature does not make disks with higher surface brightnesses),
rather than a selection effect propagating from an optical surface
brightness cutoff in the $B$ band.  If it were the latter, then we
would not expect the surface brightness of the cutoff to be constant
with the near-IR absolute magnitude, given the strong dependence of
the optical to near-IR color with galaxy luminosity.  This physical
cutoff further supports our belief that our sample does contain
examples of galaxies with normal disk surface brightnesses, in
addition to examples with disk surface brightnesses over 3.5
magnitudes fainter, thus giving us a true continuum of galaxies from
normal disks to extremely low surface brightness systems.

Third, we have also included in Figure \ref{MJ_mu0} a plot of the
dependence of the bulge central surface brightness on the near-IR
absolute magnitude.  Unlike for the disks, we do see a correlation
when {\emph{overall}} surface brightness is considered (with a
Spearman test suggesting that less than a 10\% chance that they are
uncorrelated for the low HI mass subsample).  In addition, while
Figure \ref{MJ_mu0} shows a cutoff at bright surface brightnesses for
the disk component, the bulges in these same galaxies extend to much
higher surface brightnesses (with a possible, but less sharp cutoff at
$\mu_J(bulge)\sim 16$ mag arcsec$^{-2}$).  For a comparison with the
optical see Figure 4a from \citet{oneil2000}.

\placefigure{MJ_mu0}

Fourth, we compare the sizes (in kpc) and surface brightnesses of the
bulge (left) and disk (right) components in Figure
\ref{lsb_h_mu0_bd_J}. It is apparent that LSBGs of the high HI mass
subsample B (open circles) have brighter bulge surface brightnesses
but similar bulge sizes, compared with LSBGs of the low HI mass
subsample A (filled circles).  A different trend, however, is observed
for disks (right panel). In this case, galaxies of sample A have
systematically smaller disk scale lengths with respect to those of
sample B, but with little change in typical surface brightness.
Therefore, galaxies of sample B have large disks and very high surface
brightness bulges.  Those galaxies with lower HI masses have small
disks, and bulges that are structurally similar to those of the high
HI mass sample, but that have very low surface brightness.  The same
behavior can be seen in the $K_s$ band (however we have better
signal-to-noise in $J$).  

\placefigure{lsb_h_mu0_bd_J}

Rather than just comparing our two samples against each other, we can
also compare morphology with HI content directly, by examining the bulge
and disk structural parameters derived above as a function of HI mass.
In Figure \ref{lsb_mass_mu0_J} the relationship is shown for the bulge
and for the disk central surface brightness, in the left and the right
hand panels, respectively.  It is clear that the galaxies with low HI
masses are almost entirely lacking in high surface brightness bulges,
leading to a much smaller dispersion among the bulge central surface
brightnesses.  In contrast to the trends in bulge central surface
brightness, there are no strong trends seen in the disk surface
brightnesses.  Instead, there is a very large dispersion among the
disk properties, independently of the gas mass of the galaxies.

\placefigure{lsb_mass_mu0_J}

Note that in all the above analyses using the near-IR bands, we are
sampling nearly {\em the same} stellar populations both in the bulge and in
the disk, namely red giants and old stellar populations.  Thus, the
relative sizes and surface brightnesses reflect true differences in
the distribution of stellar density.  Note also that these
measurements of the current stellar content are not {\emph{a priori}}
related to the remaining HI content after star formation.  However,
the strong trends seen in Figure \ref{lsb_h_mu0_bd_J} and Figure
\ref{lsb_mass_mu0_J} strongly suggest that the stellar surface density of a
galaxy is indeed tightly coupled to the amount of neutral gas the
galaxy currently has, though probably indirectly through the total
galaxy mass, as we now demonstrate below.

\subsection{Comparison between HI content and near-IR properties}

In the above analyses we have begun to consider how some properties
vary as a function of the total HI mass.  However, it is possible (if
not likely) that the HI mass is not the sole controlling quantity
driving these trends.  It is quite possible that the HI mass
(M$_{HI}$) correlates closely with the overall galaxy mass --
i.e. that ``bigger'' galaxies have more of everything.

In order to begin examining this possibility, Figure \ref{mass_mag}
shows the strong correlation we observe between the HI mass and the near-IR
absolute magnitudes.  Indeed, to first order, galaxies which have more
stars (as measured by the near-IR luminosity) have more HI.  Note that
there are $\sim 8$ magnitudes between the most HI massive galaxies and
the less massive ones.  Recall the striking fact that this correlation
(which is well defined in both $J$ and $K_s$) is between the mass of
the {\em HI gas}, and the near-IR luminosity of the {\em stellar
populations}; {\emph{a priori}} there should not be a correlation
between the mass in stars a galaxy has formed and the amount of gas
it has left.

In Figure \ref{mass_mag}, straight lines represent constant values of
the ratio of gas mass to stellar luminosity
for M$_{HI}$/L = 0.01 (long-dashed line), 0.1 (solid line), 1.0
(dotted line), and 10.0 (dashed line). Note that most LSBGs of our
total sample have M$_{HI}$/L $\la$ 1.0 both in $J$ and $K_s$, with an
{\em average} of M$_{HI}$/L $\sim 0.1$ and large scatter.  While the
scatter is over an order of magnitude in relative proportion of stars
and gas, it is much smaller than the overall range of 8 orders of
magnitude in overall gas mass, and thus does not wipe out the
correlation between HI mass and total stellar luminosity.  The overall
trend in Figure~\ref{mass_mag} suggests that the division between our
samples A and B is largely one mass, albeit with a wide range of gas mass to
stellar luminosity ratios.

\placefigure{mass_mag}
\placefigure{M_mlratio}

We may further examine our sample for trends in how the gas mass to
stellar luminosity ratio varies as a function of the overall galaxy
luminosity.  Figure \ref{M_mlratio} shows the logarithmic gas mass to
stellar luminosity ratio
as a function of $J$ and $K_s$ absolute magnitudes (where we have
adopted $M_\sun (J) = 4.03$ and $M_\sun (K_s) = 3.33$).  Although
there is considerable scatter, the lower luminosity galaxies have a
higher fractional gas content, on average.  The straight lines in
Figure \ref{M_mlratio} are least square fits, giving slopes of $2.86
\pm 0.46$ in $J$ and $3.35 \pm 0.44$ in $K_s$.

The gas mass to stellar luminosity ratio as a function of the absolute
magnitude (Figure \ref{M_mlratio}) indicates the relative importance
of gas and stars in the total baryonic mass budget (albeit
indirectly). Galaxies with high $M_{HI}/L$ in the near-IR have been
historically less efficient at converting their gas into stars, due to
either young mean age (i.e.\ a late start), low star-formation
efficiency, large late-time gas infall, fewer past episodes of
HI-stripping, or some combination of the above.  The fact that we see
some correlation between $M_{HI}/L$ and the galaxies' luminosities
suggest that overall there are indeed trends between galaxies' star
formation histories and their stellar mass.  However, the large
scatter suggests that it is not a perfect one-to-one relationship.  We
will address these issues in more detail in subsequent sections, where
we attempt to reduce the scatter by calculating the actual stellar
mass, rather than using the near-IR luminosity as a substitute.

Another possible quantity which correlates with M$_{HI}$ is the galaxy
morphology.  The study of morphological types shows a modest
correlation with HI content.  Within our sample, a set of LSBGs with
similar HI masses have morphological types spanning the entire Hubble
sequence from dIm to Sa.  However, despite the large scatter, a mean
trend does exist for morphological types over the whole HI mass
interval.  Galaxies of sample A (with low HI masses and, via
Figure~\ref{mass_mag}, low stellar luminosities) exhibit morphological
types of dIm, Sm, Sd, and only a few earlier type Sc and Sb. Only
three cases are Sa. In contrast, LSBGs of sample B are much earlier
type, and are classified in general as Sc, Sb and Sa, although some
cases of Sm are also present (see Figures \ref{mosaics1} and
\ref{mosaics2}, and text below).

Much stronger trends are seen between the morphology of the galaxies
and their gas mass to stellar luminosity ratio.  Figures
\ref{sort1_ml} and \ref{sort2_ml} show images of the galaxies sorted
by their M$_{HI}$/L$_{K_s}$ values.  The range in $M_{HI}/L_{K_s}$
goes from M$_{HI}$/L$_{K_s}$ = 0.028 - 0.128 $M_\sun/L_\sun$ in
Figure~\ref{sort1_ml} (from LSBG \#036 with log($M_{HI}/M_\sun$) =
8.74, to LSBG \# 437 with log($M_{HI}/M_\sun$) = 7.92), and continues
up to M$_{HI}$/L$_{K_s}$ = 1.475 $M_\sun/L_\sun$ in
Figure~\ref{sort2_ml} (galaxy LSBG \#446 with log($M_{HI}/M_\sun$) =
9.63).  It is apparent in these two Figures that as M$_{HI}$/L$_{K_s}$
becomes larger, galaxies extend to much later Hubble types, beyond the
text-book Hubble sequence.  Galaxies with small M$_{HI}$/L$_{K_s}$ are
earlier Sa/Sb/Sc types, while galaxies with large $M_{HI}/L_{K_s}$ tend
to be Irr galaxies, suffer from tidal distortions, plumes and other
peculiarities.

\placefigure{sort1_ml} \placefigure{sort2_ml}

\subsection{Stellar mass-luminosity ratios}

In previous sections we have considered trends as a function of the HI
mass to stellar luminosity. This ratio gives a measure of the
evolutionary state of the galaxies, indicating how far the galaxy has
proceeded along the path of converting gas into stars.  However, the
infrared stellar luminosity is only an approximate indicator of the
stellar mass.  We may obtain clearer and more physically meaningful
results by using stellar population synthesis codes to translate the
stellar luminosities into estimates of the actual stellar masses,
allowing us to make a direct comparison between the gaseous and
stellar mass.

To convert from IR luminosity to stellar mass, we may take advantage
of the work of \citet{bell2001}, which demonstrated that although the
stellar mass-to-light ratio varies by up to a factor of 2 in the
near-IR, there are clear trends between the stellar M/L and color.
Using spectrophotometric spiral galaxy evolution codes,
\citet{bell2001} showed that one can use the colors of galaxies to
derive their relative stellar mass-to-light ratios (under the
assumption of a universal spiral galaxy initial mass function).  They
also showed that uncertainties in the dust-reddening estimates do not
strongly affect the final derived stellar masses of a stellar
population.

For testing whether the variations of M/L with $J - K_s$ color are
robust to different stellar population synthesis models, we have used
\citet{bell2001}'s relations between the stellar M/L and the $V - J$
and $V - K$ colors, to derive the relationship between M/L and our $J
- K_s$ color\footnote{Differences between $K$ and $K_s$ are less than
$\sim 0.08$ mags.}. Three different models are tested, which are (1)
an instantaneous burst with a Scalo IMF, from recent Bruzual \&
Charlot models \citep{liu2000}, (2) a recent version of
PEGASE\footnote{PEGASE is ``Projet d'Etude des GAlaxies par Sinth\`ese
Evolutive.} models \citep{fioc97,fioc2001} with an IMF slope of
$-1.85$, and a model with a formation epoch with bursts, adopting a
Salpeter IMF. All models have solar metallicity. The key point is that
M/L vs color slopes do not depend on the IMF, which only affects the
zero point.  The stellar masses do not vary more than 7\% between the
two models. We have also used sub-solar metallicities and again the
variation in the stellar masses are $\sim 10\%$, and consistent with
the variations between models. In what follows, we adopt the Salpeter
model (using Table 1 from \citet{bell2001}) and obtain coefficients for
M/L as a function of $J - K_s$ color\footnote{\citet{bell2001} note
that these relations have been calibrated for maximal disks, and may
be slight overestimates if the disks are submaximal.}.  With the
adopted relationship between stellar mass to light ratio and $J-K_s$
color, we can compute the {\em actual} baryonic masses $M_{baryonic}$
from the derived stellar masses $M_{stellar}$
($=L_{K_s}\times(M/L_K)$) and the HI mass as $M_{baryonic} =
M_{stellar} + M_{HI}$.  We are implicitly assuming that the mass
contribution from molecular, ionized, and metallic gas is negligible for our
sample.

\placefigure{Mstars_relations}

Figure \ref{Mstars_relations} shows the gas-to-star mass fraction as a
function of stellar mass, after computing the stellar masses as
described above.  Figure \ref{Mstars_relations} clearly shows that
LSBGs with larger stellar masses tend to have smaller gas fractions,
and LSBGs with small star masses have larger gas fractions.

We have also included in Figure \ref{Mstars_relations} diagonal lines
representing constant HI mass; dashed line for $M_{HI} = 7.0$ M$_\sun$
(close to our lowest HI mass), solid line for 9.0 M$_\sun$ (upper
limit of our sample A), dotted line for 9.5 M$_\sun$ (lower limit for
our sample B), and long-dashed line for 10.0 M$_\sun$ (close to our largest HI
mass).  These lines can be used for assessing the importance of
selection criteria.
For example, there seems to be a lower limit for the mass fraction
HI-to-stars at $\sim 5 \times 10^{-2}$.  However, these LSBGs can
exist but their HI would not be detectable, and thus they may not be
present in the subsample of galaxies with measured HI fluxes that we
selected from the \citet{impey96} catalogue.

Figure \ref{Mstars_relations} also confirms our conclusions (1) that
low HI mass does not imply that a galaxy has ``used up'' its gas
reservoir and converted it entirely to stars (given the existence of
galaxies with low $M_{HI}$ and high $M_{HI}/M_{star}$) and (2) that
likewise high HI mass does not imply that a galaxy has only begun to
convert HI into stars (given the large numbers of galaxies with large
HI masses but large stellar masses as well).  In other words, both
high HI mass galaxies {\em and} low HI mass galaxies have a large
range of stellar masses (with low HI mass galaxies having stellar
masses between $10^{7}$ and $10^{10}$ M$_\sun$, and high HI mass
galaxies having stellar masses between $10^{9}$ and $10^{12}$
M$_\sun$), and thus selecting on HI alone is not sufficient to identify
galaxies which are gas-rich or gas-poor.

We have also plotted similar relations as in
Figure~\ref{Mstars_relations}, but using the ratio of the HI mass to
the total baryonic mass ($M_{stars}+M_{HI}$), shown in Figure
\ref{Mbaryonic_relations}.  Note that the ratio
M$_{HI}$/M$_{baryonic}$ has an upper limit, when all the baryonic mass
is concentrated in the gas (ratio = 1).  We see again that increasing
baryonic mass implies that a galaxy has converted a much larger
fraction of its baryonic mass into stars.  We have also included in
Figure \ref{Mbaryonic_relations} the locus separating galaxies
according to their dominant component, stars or gas.  Again, note that
``star-dominated'' LSBGs can still have large quantities of HI, and
similarly, ``HI-dominated'' LSBGs can have a small HI-mass if the
overall baryonic mass is small.  The extreme galaxy at M$_{baryonic}
\sim 3 \times 10^{11}$ M$_\sun$ is LSBG 224 which, even with a large
HI content ($M_{HI} = 8.0 \times 10^{10} M_\sun$), has a baryonic mass
represented almost exclusively by the stars. It is clearly a prominent
Sa galaxy (see Figure \ref{sort1_ml}).  In general, all of the
star-dominated LSBGs are Sa/Sb galaxies, whereas the HI-dominated
LSBGs have morphologies ranging from spirals to distorted or fuzzy
irregulars.  Among the most gas rich systems we find the most
irregular and fuzzy LSBGs (HI-rich galaxies) in our sample. Note that,
although the number of galaxies from sample A and B is roughly similar,
most of the galaxies lie below the dotted horizontal line, and therefore 
are star dominated ones.

\placefigure{Mbaryonic_relations}

If we investigate these morphological trends further, we discover that
both the stellar mass and the stellar-to-HI mass fraction are related
closely to surface brightness. Figure \ref{mu0_relations} shows the
bulge central surface brightness as a function of the ratio
M$_{HI}$/M$_{baryonic}$, and as a function of the total baryonic mass.
The ratio M$_{HI}$/M$_{baryonic}$ increases with decreasing surface
brightness, showing that at higher bulge surface brightnesses,
galaxies have produced a much larger fraction of stars over their
lifetime.  It may be that the formation of a bulge creates an
extremely efficient epoch of star formation, as gas is driven to very
high central densities.  Given the observed correlations between gas
density and star formation rate \citep{kennicutt98}, then we would
expect that galaxies with the highest stellar densities (i.e. the high
surface brightness bulges) had the highest past star formation rates,
and thus the lowest current gas fractions. In other words, all
galaxies are ``born'' at the locus of M$_{HI}$/M$_{baryonic}\!=\!1$,
and then travel downwards and to the left in
Figure~\ref{mu0_relations} to M$_{HI}$/M$_{baryonic}\!=\!0$, at a rate
which depends in large part upon the central surface density of the
gas. The end result is a lack of high surface brightness
galaxies with large gas mass fractions, and low surface brightness
galaxies with low gas mass fractions (e.g.\ the region around $\mu_0
\sim 25$ mag arcsec$^{-2}$ and M$_{HI}$/M$_{baryonic} \sim 0.1$).

\placefigure{mu0_relations}

We may explore these ideas more directly by converting the observed
surface brightnesses to stellar surface densities, using the
color-dependent stellar mass to light ratios calculated above.  In
Figure \ref{stellar_density} we have plotted the $J - K_s$ color and
gas to baryonic mass ratio as a function of the central stellar
surface density of the bulge $\Sigma$ (in M$_\sun$/pc$^2$).  We see a
very strong trend (correlation coefficient of 0.22) towards lower gas
fractions with increasing central surface densities, as expected for
increased star formation efficiency at higher surface densities.  The
scatter is also reduced over Figure \ref{mu0_relations}.  In addition,
we see a trend towards redder colors with increasing surface density,
but we will defer a discussion of this effect until \S\ref{colorsec} below.

\placefigure{stellar_density}

While the increase in M$_{HI}$/M$_{baryonic}$ is a steady function of
the bulge surface brightness, an equally striking trend is found with the
the total baryonic mass.  Figure~\ref{mu0_relations} shows that the
formation of a high surface brightness bulge is a very sharp function
of the overall baryonic mass.  Galaxies with baryonic masses of less
than M$_{baryonic}\!=\!10^{10}\msun$ simply do not form bulges (using
$\mu_0(bulge) \la 18.0$ mag arcsec$^{-2}$, marked by a dash-dotted
vertical line in Figure~\ref{mu0_relations}, as the demarcation of
galaxies which do not have bulges significant enough to be detectable
in the near-IR, i.e.\ those which cannot be distinguished clearly from
the disk)\footnote{For the galaxies with $\mu_0(bulge) \ga 18.0$
mag arcsec$^{-2}$, we find mostly irregular morphologies, with
a larger tendency toward showing spiral morphologies when dominated
by stars (i.e.\ low M$_{HI}$/M$_{stars}$).}.

The floor in the baryonic mass suggests that there is a tight coupling
between the mass of the galaxy and the conditions necessary for bulge
formation.  We can envision such a correlation arising as follows.
Assuming that there is no large scale segregation between baryons and
dark matter, then the baryonic mass of the galaxy should be directly
proportional to the total mass of the galaxy
(M$_{total}$/M$_{baryon}\sim16$, for $\Omega_m=0.25$ and
$\Omega_{baryon}=0.015$).  Because the total mass of a galaxy is
closely coupled to several astrophysical timescales for galaxy
formation (such as the epoch of first collapse, or the
end of an epoch of merging), one might expect to see a correlation
between {\emph{tracers}} of these quantities and the baryonic mass.
In many scenarios, the bulge forms via a by-product of early merging in
the galaxy, or via the collapse of low angular momentum material in the
galaxy -- both of which probably depend closely on the above
timescales.  Thus, it may be that future models of galaxy formation
can use this observed floor for bulge formation as a critical test.

The final question arises as to whether or not selection effects play
a role in shaping the trends in Figure~\ref{mu0_relations}.  For
example, the lack of galaxies with high surface brightness bulges in
the near-IR and low baryonic masses (i.e. the dashed oval) may be due
to the fact that our sample was drawn from a catalog of low surface
brightness galaxies.  Such a catalog would be biased against bright
bulges, unless accompanied by a low surface brightness disk, which
would increase the overall mass of the galaxy, possibly pushing it up
above the $10^{10}\msun$ floor.  However, we suspect that the floor in
Figure~\ref{mu0_relations} is real, for several reasons.  First of
all, the existence of large numbers of galaxies with $\mu_0(bulge) \ga
18$ mag arcsec$^{-2}$ suggests that the sample by no means excludes
galaxies of high near-IR surface brightness galaxies.  Indeed, an
analysis by \citet{sprayberry96} shows that the original
\citet{impey96} sample includes a large fraction of high
{\emph{optical}} surface brightness galaxies, indistinguishable from
``normal'' spirals.  Second, if the addition of a low surface
brightness disk were necessary for inclusion in the original Impey
catalog, there is no reason why the lower left of the plot should not be
populated with low mass galaxies that would otherwise have dropped off
the bottom of the plot.  Finally, the observational selection effects
would be most apparent in a plot of purely observational quantities
(for example, $M_J$ vs $\mu_0(bulge,J)$), but in such a plot, we see
the plane is evenly populated (e.g.\ Figure~\ref{MJ_mu0}), and we can
identify no steps in the process of converting from $M_J$ to
M$_{baryonic}$ that would lead to the creation of an artificial floor
in the baryonic mass associated with high surface brightness bulges.

We stress that the fact that no gas poor but low surface brightness
galaxies with very low values of M$_{HI}$/M$_{baryonic}$ and faint
surface brightness appear in the Figure ($\mu_0 \sim 25$ mag
arcsec$^{-2}$ and M$_{HI}$/M$_{baryonic} \sim 0.1$), which could be a
combined effect of our initial selection on surface brightness and gas
content ($\mu_0(B) \le 23.5$ mag arcsec$^{-2}$), even though we have
not imposed a low cut for the HI mass. We know that large galaxies
with bright total magnitudes but extremely low surface brightness and
large disk scale lengths do exist, like Malin 1 (with $M_B \sim
-23.0$, $\mu_B = 26.5$ mag arcsec$^{-2}$ and $M_{HI} \sim 10^{11}
M_\sun$, \citet{matthews2001}) and are placed at such a locus in the
diagram, where spirals with faint bulges and large stellar masses lie.
However, Figure \ref{mu0_relations} suggests that these galaxies are
rare.

\subsection{The interplay between color and mass content} \label{colorsec}

In the previous section we have explored trends involving
M$_{HI}$/M$_{baryonic}$ as an indicator of the evolutionary state of a
galaxy.  If a galaxy steadily transforms its initial reservoir of gas
into stars, then as M$_{HI}$/M$_{baryonic}$ decreases, the mean
metallicity of the stellar population should increase as well.
In this section we explore metallicity trends within our sample,
after demonstrating that the near-IR $J-K_s$ color is an excellent 
indicator of the mean stellar metallicity.

Within optical wavelengths, the variations in the broad band colors of
galaxies are thought to be primarily driven by changes either in the mean age
of the stellar population or in the dust content \citep{bell2001}.
However, in the near-IR the situation is quite different.  First of
all, reddening due to dust is negligible in all but the most extreme
starburst galaxies. Secondly, and most importantly for this analysis,
changes in the near-IR color of a galaxy are driven almost entirely by
the mean {\emph{metallicity}} of the galaxy; in the near-IR, the light
from a galaxy is dominated by stars on the red giant branch, the
location of which shifts to redder or bluer colors depending on
metallicity.  To demonstrate this effect, in Figure \ref{gridplot} we
plot stellar population synthesis models from Bruzual \& Charlot
\citep{liu2000} of the $J - K_s$ color as a function of $B - V$, for
different metallicities and exponentially declining or increasing
star-formation rates (i.e.  with different mean stellar ages),
assuming that star-formation started 12 Gyr ago.  The $J - K_s$ color
index is clearly {\em very} metal sensitive, and relatively age
insensitive; see \citet{bell2000}.  At a single stellar metallicity,
the change in $J-K_s$ with mean stellar age is never more than
$\sim\!0.15$, even when the mean age of the stellar population changes
by 11 Gyr.  This implies that whenever one witnesses a change of
more than $\Delta(J-K_s)\sim0.15$ then one must be witnessing
a significant change in stellar metallicity.  We now apply these results
to observations of the color distribution within our sample.

\placefigure{gridplot}

\placefigure{abs_mag}

In Figure \ref{abs_mag} we show color-magnitude diagrams for galaxies
appearing in Table \ref{phot1}.  The plot clearly shows that LSBGs
span a wide range in IR color ($\Delta(J-K_s)\sim0.6$), immediately
suggesting a wide range in the stellar metallicities of the galaxies.
As argued above, the largest possible shift in color due to age
changes alone ($\Delta(J-K_s)\sim0.15$) is far too small to explain
the full range of colors seen in our data.  In other words, although
changes in mean stellar age may contribute to the range in color, they
cannot be responsible for all of it.  

We may place some limits on the {\emph{maximum}} range of metallicity
compatible with the observed spread in $J-K_s$ color by assuming that
the mean stellar age is constant (i.e. that all the color changes are
due to metallicity alone).  Based upon the grids in
Figure~\ref{gridplot}, a factor of 10 change in the metallicity
produces changes in color of roughly $\Delta(J-K_s)\sim0.2-0.3$.
Therefore, our observed range in galaxy colors suggests a maximum
change in mean stellar metallicity of more than a factor of 100 across the
entire sample.  Alternatively, we can calculate a {\emph{minimum}}
range in metallicity by allowing for the maximum spread in age ($\Delta(J-K_s)\sim0.15$).  The remaining color difference still implies a minimum range
of over a factor of 20 in mean stellar metallicity.  These two limiting
cases suggests that over the mass range in our sample, we have a range
in stellar metallicity covering a factor of $20-100$.

In addition to demonstrating the overall range of near-IR colors,
Figure \ref{abs_mag} shows a statistically significant trend between
the IR color and the stellar luminosity of the galaxy, suggesting a
trend between metallicity and stellar mass.  (We may also use Figure
\ref{abs_mag} to compare the typical colors of the low gas-mass and
high gas-mass subsamples that the lower HI mass galaxies in sample A
are bluer that sample B by $\sim 0.3$ magnitudes, on average,
immediately suggesting a mass dependent trend in galaxy metallicity,
given the trends between HI mass, total stellar mass, and baryonic
mass between the two samples).  We can explore these trends more
explicitly in Figure \ref{mass_JK}, where we plot $J - K_s$ directly
as a function of HI mass.  As expected there is a systematic trend
between the HI mass of a galaxy and its near-IR color.  Finally we
plot $J - K_s$ as a function of the stellar mass
and of the total baryonic mass in
Figure~\ref{color_mass}.  Again, we see trends in the near-IR color
with the stellar and baryonic masses (with a slope of $0.22 \pm 0.03$
and a rms of 0.12 for the linear fit of color versus logarithmic
stellar mass, and with a slope of $0.27 \pm 0.03$ with a rms of 0.13
for the linear fit with logarithmic total baryonic mass).

\placefigure{mass_JK}
\placefigure{color_mass}

The general picture painted by Figures~\ref{abs_mag}-\ref{color_mass}
is one where as the stellar mass, gas mass, and/or total baryonic
mass within a sample of galaxies increases, the galaxies becomes redder in
$J-K_s$, and thus more metal rich on average.  However, around all of
these relations there is substantial scatter, suggesting that none of
these quantities is the principal driver of metallicity.

\placefigure{ratio_lum_color}

Both our intuition and our data suggest that the true driver of the
metallicity is the gas mass fraction M$_{HI}$/M$_{baryonic}$, which
indicates the degree to which a galaxy has completed its lifetime of
star formation.  Galaxies that have only just begun to create stars
have only begun to enrich their gas reservoirs, and thus
should have a low mean stellar metallicity and very blue $J-K_s$
colors.  Likewise, galaxies that have completely converted their gas
into stars via several generations of star formation will have greatly
enriched the gas from which the last half of the stars formed, leading
to much more metal rich stellar populations and redder $J-K_s$ colors.
In Figure \ref{ratio_lum_color} we plot the gas mass fraction
M$_{HI}$/M$_{baryonic}$ as a function of the metal sensitive $J-K_s$
color, and overall the behavior is just as expected.  Compared to the
previous figures, there is a substantial tightening of the
relationship, especially towards the more stellar dominated systems
(i.e.\ low values of M$_{HI}$/M$_{baryonic}$).  This suggests that the
typical stellar metallicity of a galaxy is indeed largely set by the
degree to which it has already converted its gas into stars.  We have
already seen hints of this effect in Figure~\ref{stellar_density}, where
the $J-K_s$ color is highly correlated with the central surface density
of the galaxy, which in turn sets the efficiency of converting gas
into stars and thus the subsequent value of M$_{HI}$/M$_{baryonic}$.

We may also gain additional insight by considering the scatter in
Figure~\ref{ratio_lum_color}.  In a ``closed box'' model for chemical
enrichment, the mean metallicity of the stellar population increases
monotonically with decreasing gas mass fraction, leading to a perfect
one-to-one relation between the two quantities.  If the closed box
model is valid on the scale of galaxies, then the relationship between
$J-K_s$ and M$_{HI}$/M$_{baryonic}$ should have zero scatter (modulo
the small differences in $J-K_s$ produced by changes in mean age).
For the galaxies in Figure~\ref{ratio_lum_color} where star formation
has proceeded largely to completion (M$_{HI}$/M$_{baryonic}<0.2$), the
scatter is comparable to our observational errors, suggesting that
evolution could have proceeded roughly along a closed box
pathway\footnote{The gas poor galaxies are also those with the most
  substantial bulges, and thus it may be that the closed box
  approximation is valid for the bulges alone.}, as also concluded by
\citet{bell2001} for normal spirals.  In contrast,
for more gas rich galaxies, the scatter in $J-K_s$ is significantly
larger than our uncertainties, suggesting that the closed box model is
not a good approximation to the gas and metallicity evolution of these
systems.  Instead, these systems may still be in a phase of sporadic
gas infall, as opposed to the more massive star-dominated galaxies where the
gas infall seems to have run to completion.  Note also that these gas
rich galaxies span a much wider range in stellar luminosity (and thus
stellar mass) at fixed M$_{HI}$/M$_{baryonic}$, possibly suggesting a
much wider range of mean stellar age, which could contribute to the
larger scatter as well.

\placefigure{jk_mass_lum_ratio}

To facilitate comparison with earlier work in the literature, we have
also included in Figure~\ref{jk_mass_lum_ratio} plots of the $J-K_s$
color as a function of M$_{HI}$/L, which has traditionally been used
as an indicator of the relative fraction of gas and stars.  The
observed trend toward lower M$_{HI}$/L and redder colors is observed
in both bandpasses, and while it
is more obvious in the $K_s$ band, the relationship is still much
weaker overall than the trend with M$_{HI}$/M$_{baryonic}$ in
Figure~\ref{ratio_lum_color}.  We fit the trend with straight lines,
giving a slope of $-0.05 \pm 0.04$ in $J$ (which is not significant),
but a slope of $-0.13 \pm 0.04$ in $K_s$, which is. Note that
\citet{oneil2000} found {\em no relationship} between the gas fraction
and the $B-V$ color, due to the large scatter (in that work the gas
fraction and color relationship arise from a compilation of different
sources of data).  Our observed tightening of the relationship
reflects \citet{bothun1982}'s observations that near-IR colors are
less scattered than blue colors when plotted against the gas fraction.

We may explore the possible effects of age by considering previous
work on the optical colors of LSBG's.  For example, one may compare
Figure \ref{mass_JK} with Figure 5a presented by \citet{oneil2000},
where {\em no correlation} exists between the HI mass and the $B-V$
(optical) color.  The lack of any correlation between HI mass and the
age sensitive $B-V$ color index suggests a stochastic star formation
history, as also argued by \citet{schombert2001} for LSB dwarfs.  The
measured colors are luminosity weighted, and thus very small changes
in the current star formation rate produce large changes in the
luminosity-weighted mean age, and thus in $B-V$.  As quoted by
\citet{oneil2000}, this clearly suggests the existence of ``dormant''
populations of LSBGs galaxies, which can have large quantities of HI
but low current star formation.

If we take the increased scatter in Figure~\ref{ratio_lum_color}
towards gas rich galaxies as an indication of continuing gas infall,
then we may simultaneously have an explanation for the sporadic past
star formation history suggested by the optical data, namely gas
density dependent star formation thresholds.  In this scenario,
ongoing gas infall occasionally raises the gas surface density above a
critical threshold, triggering a burst of star formation and a strong
bluing of the optical colors, whiled simultaneously diluting the
metallicity of the gas phase and of subsequent generations of stars.
This would yield a larger scatter both in metallicity and in age,
leading to the much larger spread in $J-K_s$ colors.  It would also
help to explain why many of the optically faint LSBGs in our sample are
sufficiently old to have developed significant red giant branch populations,
in spite of what appear to be young luminosity weighted ages implied
by their typically blue optical colors.

Finally, we note that all of the results above are based upon the
color of the entire galaxy.  If we were to consider only the disk
component, the relationship between gas-to-star mass-luminosity ratio
and color would be loosened with the increased scatter being due to
the still low S/N which is obtained for the disk photometry in the near-IR.


\section{Conclusions}     \label{conclusionsec}

The near-IR photometry of low surface brightness galaxies presented in
this paper constitutes one of the largest database published until
now.  Because of our original selection of galaxies from the \citet{impey96}
catalog, our sample includes not just a very large number of
low surface brightness galaxies, but a number of normal spiral
galaxies as well.

The total sample studied provides good indication that an old, red
population of low surface brightness galaxies does exist.  Many of the
LSBGs analyzed exhibit clear bulges and have high surface brightness
disks in the near-IR, even though they were originally identified in a
catalog of optical LSBGs\footnote{These bulge dominated LSBGs also
tend to be red, as found by \citet{bell2000} for the optically
selected red LSBGs of \citet{oneil97}}.  Morphologically, the sample
spans the full range of spiral Hubble types, from Sa to Im.  While
there are broad correlations between the galaxies' optical and near-IR
morphologies, we find several cases of galaxies which appear to be
late-type in the optical, but have prominent bulges in the near-IR
(e.g.\ LSBGs \# 100, 338, 384, and 446 in Figures \ref{mosaics1} and
\ref{mosaics2}).

In addition to the morphologies, a quantitative analysis of the
surface brightness profiles of the galaxies suggests that our sample
contains galaxies with disk surface densities comparable to normal
spirals, as well as a very high fraction of truly low surface
brightness (and/or low surface density) disks.  On the bright end, we
find that the central surface brightnesses of the disks in our sample
show a well defined cutoff brighter than $\mu_J(disk)\sim 17.5$ mag
arcsec$^{-2}$ (independent of the $J$-band absolute magnitude).  We
argue that this cutoff is physical, and reflects the maximum surface
density at which disks are stable.  On the LSBG end, our sample
includes a very high fraction of galaxies that are indeed low surface
brightness even in the near-IR, with some reaching roughly 3.5
magnitudes fainter in surface brightness than the bright cutoff.  The
disks also have sizes that are systematically smaller for lower mass
galaxies.

As part of our analysis, we have calculated the total
{\emph{baryonic}} masses (M$_{baryonic}=M_{gas}+M_{stars}$) of
galaxies in our sample.  For M$_{gas}$, we have taken the HI mass from
the \citet{impey96} catalog, and assumed that other gaseous components
are negligible.  For M$_{stars}$, we have used the color-dependent
stellar mass-to-light ratios of \citet{bell2001} to translate the
observed near-IR luminosity into stellar mass.  The mass-to-light
ratios also allow us to convert observed surface brightnesses into
stellar mass surface densities.  The resulting masses and surface
densities allow us to consider trends involving fundamental physical
quantities (such as the ratio of the gas mass to the total baryonic
mass).  The total baryonic mass itself is probably a good indicator of
the total mass of the galaxies, assuming that baryon and dark matter
are evenly mixed on large scales.

Among the masses of the various components, we see strong
correlations, many of which follow trends previously identified in the
optical (for example, see compilations in \citet{schombert2001}).
Galaxies with large stellar masses tend to also have high HI masses,
higher central surface densities, but small gas mass fractions (i.e.\
are HI-poor).  Likewise, galaxies with small stellar masses tend to
have low HI masses but have a larger fraction of their baryonic mass
in the gas phase (are HI-rich, as for the LSBGs studied by
\citet{mcgaugh97}).  We note that even galaxies which have formed
large masses of stars can still retain large reservoirs of neutral
gas, even after rich episodes of star formation.

We find several strong morphological correlations with the masses of
the galaxies.  There are systematic trends towards more irregular,
diffuse morphologies and lower surface densities with increasing gas
mass fraction (or alternatively, M$_{HI}$/L$_{K_s}$).  We also find that
the central surface brightness of the bulge component varies strongly
with the near-IR luminosity and HI mass, as well as with the stellar
and the total baryonic mass (such that the bulges have higher surface
brightness for larger values of the listed quantities).  Moreover, we
find that our sample exhibits a ``floor'' in the baryonic mass
required for bulge formation, at $10^{10}\msun$; we have no galaxies
with lower baryonic masses that have formed high surface brightness
bulges.

We have also explored how the near-IR color varies within our sample.
In order to interpret the colors, we have used stellar population
synthesis models to argue that the $J-K_s$ color is highly metal
sensitive, such that redder colors imply higher metallicities.  We
argue that based upon the range of observed colors our sample spans a
factor of more than 20 in metallicity (but less than a factor of 100).
We find that galaxies have redder $J-K_s$ colors with increasing HI
mass, stellar mass, and baryonic mass, suggesting that more massive
galaxies have in general have more metal enriched stellar populations.

We find very tightest correlations between color and the gas mass
fraction M$_{HI}$/M$_{baryonic}$ in particular.  These correlations
are significantly tighter than previous optically determined trends,
even in the redder $V-I$ color (see compilation in Figure 6 of
\citet{schombert2001}).  In a closed box model for chemical evolution,
there should be a one-to-one relationship between the metallicity and
the gas mass fraction.  At small values of M$_{HI}$/M$_{baryonic}$,
this is essentially what we see, in that the observed scatter in the
metal-sensitive $J-K_s$ color is comparable to our observational
uncertainties.  This suggests that in the gas poor, bulge dominated
massive galaxies, chemical evolution has proceeded along the canonical
closed box pathway (possibly confined to the bulge, which dominates
the luminosity and thus the colors of the galaxy).

We find a different behavior among the gas rich, disk dominated, lower
mass galaxies.  In these systems there is large scatter around the
relationship between color and M$_{HI}$/M$_{baryonic}$.  This
suggests that not only is the enrichment of these galaxies on-going
(given their high gas mass fraction), but that it is not proceding as
expected for a ``closed box'' model of enrichment.  Instead, it is
likely that these gas rich systems are continuing to experience
episodes of gas infall.  These episodes of gas accretion first
increase the ratio of gas to stars, and then dilute the metallicity
of subsequent generations of stars.  

Episodic accretion would also explain the stochastic star formation
suggested by analyses of the optical colors of LSBGs
\citep{oneil2000,bell2001} and theoretical calculations
\citep{gerritsen98,jimenez98}.  Temporary increases above a gas
surface density threshold for star formation could dramatically
increase the star formation rate \citep{kennicutt98}, triggering
bursts of star formation and an increased rate of gas consumption.
This change in the luminosity-weighted mean stellar age would produce
additional scatter in the color, scatter which will be more dramatic
in the optical. 

Within our sample, we see strong evidence that the star formation
efficiency is indeed surface density dependent.  We find strong
correlations between the gas mass fraction M$_{HI}$/M$_{baryonic}$
and the central stellar surface density (i.e.\ M$_{stars}$/pc$^2$),
suggesting that galaxies with the highest initial gas surface density
have also had the highest efficiency in turning gas into stars.  This
link between surface density and star formation efficiency also
suggests that the young mean stellar age of LSBGs (as deduced from
luminosity-weighted optical colors) results principally from the
galaxies being slow to convert gas into stars, rather than from their
being slow to assemble (see also \citet{bell2000},
\citet{schombert2001}, \citet{deblok98}).  However, the evidence for
on-going gas accretion in lower luminosity galaxies suggests that this
second effect does play some role as well.

For future work, we intend to extend the current work with the
addition of optical colors from our on-going database of $B$ and 
$R$ photometry for the same galaxies presented here, as well as from
other bands provided by The Sloan Digital Sky Survey. 
We shall compute dynamical masses, 
and current gas phase metallicities using also future spectroscopic data. 
This will allow us to perform more critical and accurate 
tests of the ideas presented in this work.


\acknowledgements We acknowledge Las Campanas Observatory for granting
the large investment in observing time necessary for this project.  We
also thank all of the staff at Las Campanas.  We would like to
acknowledge Pablo Araya in the reduction of some images appearing in
this paper, as well as Andreas Reisenegger for comments and shaking
discussions around the subject presented here.  We also thank the
anonymous referee for many helpful suggestions that greatly improved
the paper.  Part of this research was funded by the Andes-Carnegie
fellowship C-12927, during a post-doctoral position at Las Campanas
Observatory (GG). GG thanks funding from ``Proyecto DIPUC 2001/14-E''.
GG and LI thanks ``Proyecto FONDAP'' Center for Astrophysics. LI
acknowledges ``Proyecto Puente PUC''.  JD was partially supported
by NSF grant AST-990862 and by the Alfred P. Sloan Foundation during
this work.

This research has made use of the NASA/IPAC Extragalactic Database,
which is operated by the Jet Propulsion Laboratory, California
Institute of Technology, under contract with the National Aeronautics
and Space Administration. The research has also made use of the VizieR
service from the Centre de Donne\'es Astronomiques de Strasbourg
(CDS), and the Digitized Sky Survey, which is produced at the Space
Telescope Science Institute under U.S. Government grant NAG
W-2166. The images of these surveys are based on photographic data
obtained using the Oschin Schmidt Telescope on Palomar Mountain and
the UK Schmidt Telescope. The plates were processed into the present
compressed digital form with the permission of these institutions.

%

%
\begin{deluxetable}{cccccccc}
\tabletypesize{\tiny}
\tablecaption{Average Zero points, color terms and extinction coefficients.
\label{calib}}
\tablehead{
\colhead{$<a_1>$} & \colhead{$<a_2>$} & 
\colhead{$<b_1>$} & \colhead{$<b_2>$} & \colhead{$<\xi_j>$} & 
\colhead{$<\xi_{k_s}>$} & \colhead{No. nights} &
\colhead{Stars/night}}
\startdata
(1) & (2) &(3) &(4) &(5) &(6) &(7) & (8) \\
\cutinhead{40-inch Swope telescope}
$-$5.22 (0.15)\tablenotemark{(9)} & $-$0.21 (0.23) & $-$5.98 (0.13) & 
0.12 (0.25) & 0.12 (0.05) & 0.07 (0.08) & 17 & 9 \\
\cutinhead{100-inch du Pont telescope}
$-$3.42 (0.12) & $-$0.08 (0.26) & $-$4.00 (0.14) & 0.07 (0.20) & 0.11 (0.06) &
0.09 (0.08) & 12 & 9 \\
\tablenotetext{(1)}{Average zero point for the $J$ filter.}
\tablenotetext{(2)}{Average color term for the $J$ filter.}
\tablenotetext{(3)}{Average zero point for the $K_s$ filter.}
\tablenotetext{(4)}{Average color term for the $K_s$ filter.}
\tablenotetext{(5)}{Average extinction value for the $J$ filter.}
\tablenotetext{(6)}{Average extinction value for the $K_s$ filter.}
\tablenotetext{(7)}{Number of nights considered for computing the average values.}
\tablenotetext{(8)}{Number of standard stars observed per night.}
\tablenotetext{(9)}{The standard deviation for the corresponding averages
        are included between parenthesis.}
\enddata
\end{deluxetable}


\begin{deluxetable}{rlrrrlrlllllccc}
\tabletypesize{\tiny}
\renewcommand{\arraystretch}{0.6}
\tablecaption{The sample of low surface brightness galaxies observed in the
near-IR. \label{phot1}}
\tablewidth{18cm}
\tablecolumns{15}
\tablehead{
\colhead{$\#$\tablenotemark{a}}&
\colhead{Name\tablenotemark{b}} &
\colhead{$cz$\tablenotemark{c}} &
\colhead{M$_{HI}$\tablenotemark{d}} &
\colhead{Exp. $J$\tablenotemark{e}} &
\colhead{Exp. $K_s$\tablenotemark{e}} &
\colhead{Ap.\tablenotemark{f}} &
\colhead{$J$\tablenotemark{g}} &
\colhead{$\Delta J$\tablenotemark{h}} &
\colhead{$K_s$\tablenotemark{g}} &
\colhead{$\Delta K_s$\tablenotemark{h}} &
\colhead{M$_J$\tablenotemark{i}} &
\colhead{M$_{K_s}$\tablenotemark{i}} &
\colhead{$J-K_s$} &
\colhead{$\Delta (J-K_s)$}}
\startdata
\multicolumn{13}{c}{low HI mass sample (sample A)} \\
\tableline
   16 & 0027+0134  & 3436  &     8.65  & 2250.0  & 2750.0  & 20.0  & 13.119   &   0.03  &   12.817  &   0.06  & -20.186 & -20.488 &  0.302  &   0.07 \\
   33 & 0101+0304  & 4379  &     8.92  & 3600.0  & 1250.0  & 22.0  & 14.742   &   0.04  &   13.901  &   0.07  & -19.089 & -19.930 &  0.841  &   0.08 \\
   36 & 0104+0140  & 4708  &     8.74  & 1800.0  & 2500.0  & 12.0  & 12.706   &   0.03  &   11.858  &   0.04  & -21.282 & -22.130 &  0.848  &   0.05 \\
   38 & 0105+0047  &  632  &     7.48  & 2250.0  & 3750.0  & 20.0  & 13.573   &   0.03  &   13.220  &   0.07  & -16.055 & -16.408 &  0.353  &   0.08  \\
  295 & 0124-0043  &  963  &     8.10  & 2760.0  & 3750.0  & 20.0  & 13.513   &   0.02  &   12.832  &   0.04  & -17.029 & -17.710 &  0.681  &   0.04 \\
  100 & 0233+0012  & 2615  &     8.89  & 2250.0  & 4250.0  & 22.0  & 13.362   &   0.02  &   12.730  &   0.04  & -19.350 & -19.982 &  0.632  &   0.04 \\
  146 & 0336+0212  & 3144  &     8.62  & 3600.0  & 5500.0  & 22.0  & 14.362   &   0.05  &   13.901  &   0.08  & -18.750 & -19.211 &  0.461  &   0.09  \\
  200 & 0918-0028  & 3492  &     8.63  & 2250.0  & 3600.0  & 19.2  & 12.622   &   0.03  &   11.804  &   0.03  & -20.718 & -21.536 &  0.818  &   0.04 \\
  201 & 0918+0147  & 4978  &     8.88  & 3000.0  & 3750.0  & 18.0  & 15.854   &   0.06  &   14.966  &   0.09  & -18.256 & -19.144 &  0.888  &   0.10 \\
  249 & 1035+0238  & 5637  &     8.89  & 3000.0  & 4000.0  & 10.0  & 16.088   &   0.06  &   15.536  &   0.09  & -18.291 & -18.843 &  0.552  &   0.10 \\
  253 & 1036+0158  &  710  &     7.62  & 3150.0  & 3500.0  & 22.0  & 13.118   &   0.03  &   12.620  &   0.05  & -16.763 & -17.261 &  0.498  &   0.06 \\
  264 & 1043+0202  & 1018  &     7.73  & 2880.0  & 4450.0  &  8.0  & 15.898   &   0.04  &   15.151  &   0.08  & -14.765 & -15.512 &  0.747  &   0.09 \\
  266 & 1047+0131  & 1604  &     7.94  & 3000.0  & 3750.0  & 13.9  & 15.653   &   0.05  &   14.910  &   0.07  & -15.997 & -16.740 &  0.743  &   0.09 \\
  270 & 1050+0253  & 1037  &     8.02  & 3150.0  & 4000.0  & 17.0  & 14.504   &   0.04  &   14.036  &   0.07  & -16.199 & -16.667 &  0.468  &   0.08 \\
  285 & 1108+0121  &  993  &     8.00  & 3000.0  & 3950.0  & 26.1  & 14.466   &   0.03  &   13.899  &   0.06  & -16.143 & -16.710 &  0.567  &   0.07 \\
  308 & 1156+0254  & 3233  &     8.72  & 4800.0  & 5500.0  & 17.0  & 15.791   &   0.04  &   15.098  &   0.07  & -17.381 & -18.074 &  0.693  &   0.08 \\
  312 & 1158-0101  & 1533  &     8.45  & 1800.0  & 2750.0  & 20.0  & 12.253   &   0.02  &   11.419  &   0.05  & -19.299 & -20.133 &  0.834  &   0.05 \\
  314 & 1205+0058  & 5870  &     8.97  & 2000.0  & 2750.0  & 21.0  & 14.034   &   0.04  &   13.156  &   0.06  & -20.433 & -21.311 &  0.878  &   0.07 \\
  319 & 1209+0305  &  883  &     7.94  & 3360.0  & 4000.0  & 15.0  & 15.350   &   0.05  &   15.008  &   0.08  & -15.004 & -15.346 &  0.342  &   0.09 \\
  329 & 1216+0029  &  867  &     7.55  & 3000.0  & 3750.0  & 70.0  & 15.411   &   0.05  &   14.940  &   0.07  & -14.903 & -15.374 &  0.471  &   0.09 \\
  330 & 1217+0103  & 2056  &     8.62  & 2400.0  & 2750.0  & 20.0  & 13.392   &   0.03  &   12.656  &   0.05  & -18.797 & -19.533 &  0.736  &   0.06 \\
  336 & 1223+0117  & 1473  &     7.59  & 3000.0  & 3750.0  & 15.0  & 14.788   &   0.03  &   14.101  &   0.06  & -16.677 & -17.364 &  0.687  &   0.07 \\
  338 & 1223-0058  & 2018  &     8.32  & 3120.0  & 3750.0  & 10.4  & 16.128   &   0.07  &   15.334  &   0.08  & -16.021 & -16.815 &  0.794  &   0.10 \\
  348 & 1227+0254  & 1635  &     8.36  & 2400.0  & 3750.0  & 15.0  & 14.074   &   0.06  &   13.612  &   0.07  & -17.618 & -18.080 &  0.462  &   0.09 \\
  349 & 1228+0157  & 1105  &     7.75  & 3000.0  & 3650.0  & 20.9  & 14.421   &   0.06  &   13.611  &   0.07  & -16.420 & -17.230 &  0.810  &   0.09 \\
  365 & 1257+0219  &  874  &     7.61  & 3000.0  & 4000.0  & 18.0  & 13.588   &   0.04  &   12.604  &   0.05  & -16.744 & -17.728 &  0.984  &   0.06 \\
  384 & 1326+0109  & 3267  &     8.74  & 3000.0  & 3750.0  & 25.0  & 14.313   &   0.03  &   13.625  &   0.06  & -18.882 & -19.570 &  0.688  &   0.07 \\
  385 & 1327+0148  & 1049  &     7.49  & 2400.0  & 3650.0  & 20.8  & 14.285   &   0.03  &   13.734  &   0.06  & -16.443 & -16.994 &  0.551  &   0.07 \\
  393 & 1350+0022  & 4605  &     8.84  & 3150.0  & 3500.0  & 18.0  & 13.797   &   0.04  &   13.190  &   0.05  & -20.143 & -20.750 &  0.607  &   0.06 \\
  398 & 1353+020   & 4738  &     8.96  & 3000.0  & 3750.0  & 17.5  & 15.088   &   0.06  &   14.706  &   0.06  & -18.914 & -19.296 &  0.382  &   0.08 \\
  400 & 1357-0017  & 4255  &     8.74  & 4800.0  & 5500.0  & 13.0  & 16.210   &   0.06  &   15.664  &   0.09  & -17.559 & -18.105 &  0.546  &   0.10 \\
  424 & 1433+0249  & 1557  &     8.03  & 4800.0  & 5250.0  & 17.0  & 15.273   &   0.05  &   14.767  &   0.08  & -16.313 & -16.819 &  0.506  &   0.09 \\
  433 & 1438+0049  & 2837  &     8.30  & 3000.0  & 5500.0  & 20.0  & 14.463   &   0.04  &   14.077  &   0.08  & -18.426 & -18.812 &  0.386  &   0.09 \\
  435 & 1439+0053  & 1891  &     8.53  & 1680.0  & 2250.0  & 27.8  & 13.291   &   0.02  &   12.661  &   0.06  & -18.717 & -19.347 &  0.630  &   0.06 \\
  437 & 1440-0010  & 1744  &     7.92  & 2400.0  & 3500.0  & 20.8  & 14.349   &   0.04  &   13.775  &   0.06  & -17.483 & -18.057 &  0.574  &   0.07 \\
  436 & 1440-0008  & 1703  &     8.46  & 2250.0  & 3000.0  & 28.0  & 12.639   &   0.03  &   12.100  &   0.07  & -19.141 & -19.680 &  0.539  &   0.08 \\
  468 & 2311-000   & 4392  &     8.95  & 1800.0  & 2750.0  & 20.0  & 12.364   &   0.03  &   11.721  &   0.05  & -21.474 & -22.117 &  0.643  &   0.06 \\
  483 & 2319+0112  & 8816  &     9.73  & 2700.0  & 4000.0  & 20.0  & 13.810   &   0.04  &   13.240  &   0.07  & -21.541 & -22.111 &  0.570  &   0.08 \\
  488 & 2327-0007  & 5207  &     8.99  & 2700.0  & 3750.0  & 30.0  & 12.776   &   0.03  &   12.145  &   0.06  & -21.431 & -22.062 &  0.631  &   0.07 \\
  492 & 2329+0203  & 5176  &     8.99  & 2700.0  & 3750.0  & 20.0  & 13.831   &   0.03  &   13.168  &   0.06  & -20.363 & -21.026 &  0.663  &   0.07 \\
\tableline
\multicolumn{13}{c}{High HI mass sample (sample B)} \\
\tableline
    4 & 0013-0034 & 11806  &    10.18  & 1650.0  & 2750.0  & 20.0  & 12.625   &   0.03  &   11.907  &   0.06  & -23.360 & -24.078 &  0.718  &   0.07  \\
   23 & 0050+0005 & 10283  &     9.83  &  800.0  & 3000.0  & 13.0  & 14.300   &   0.03  &   13.540  &   0.06  & -21.385 & -22.145 &  0.760  &   0.07 \\
   30 & 0056+0019 &  5437  &     9.62  & 1800.0  & 2750.0  & 23.0  & 11.915   &   0.02  &   11.169  &   0.05  & -22.386 & -23.132 &  0.746  &   0.05 \\
   59 & 0121+0128 &  5028  &     9.64  & 1350.0  & 1500.0  & 25.0  & 10.726   &   0.02  &    9.985  &   0.05  & -23.405 & -24.146 &  0.741  &   0.05 \\
  126 & 0311+0241 &  9533  &     9.57  & 2250.0  & 2750.0  & 12.0  & 13.986   &   0.04  &   13.238  &   0.07  & -21.534 & -22.282 &  0.748  &   0.08 \\
  165 & 0350+0041 & 11362  &     9.60  & 2250.0  & 3750.0  & 20.0  & 13.156   &   0.03  &   12.298  &   0.05  & -22.746 & -23.604 &  0.858  &   0.06 \\
  196 & 0913+0054 & 11401  &     9.68  & 2700.0  & 2750.0  & 21.0  & 13.050   &   0.04  &   12.096  &   0.05  & -22.859 & -23.813 &  0.954  &   0.06 \\
  207 & 0929+0147 & 17324  &    10.03  & 1800.0  & 2750.0  & 27.0  & 13.855   &   0.05  &   12.105  &   0.06  & -22.962 & -24.712 &  1.750  &   0.08 \\
  213 & 0954+020  &  9592  &     9.59  & 1800.0  & 2750.0  & 24.0  & 13.686   &   0.05  &   12.768  &   0.06  & -21.848 & -22.766 &  0.918  &   0.08 \\
  224 & 1007+0121 & 29213  &    10.69  & 1800.0  & 2750.0  & 24.0  & 13.217   &   0.03  &   12.119  &   0.07  & -24.735 & -25.833 &  1.098  &   0.08 \\
  225 & 1008+0128 &  9915  &     9.87  & 1800.0  & 2750.0  & 24.0  & 13.080   &   0.05  &   12.167  &   0.05  & -22.526 & -23.439 &  0.913  &   0.07 \\
  230 & 1015+0148 & 13660  &     9.84  & 3000.0  & 3750.0  & 12.0  & 14.965   &   0.03  &   13.913  &   0.04  & -21.336 & -22.388 &  1.052  &   0.05 \\
  242 & 1030+0252 &  8825  &     9.65  & 1800.0  & 2750.0  & 16.0  & 12.581   &   0.04  &   11.650  &   0.04  & -22.772 & -23.703 &  0.931  &   0.06 \\
  262 & 1043+0018 & 13966  &     9.50  & 1800.0  & 2750.0  & 10.0  & 13.911   &   0.03  &   13.118  &   0.06  & -22.439 & -23.232 &  0.793  &   0.07 \\
  265 & 1045+0014 & 11644  &     9.57  & 1800.0  & 2500.0  & 12.0  & 13.747   &   0.05  &   12.898  &   0.06  & -22.208 & -23.057 &  0.849  &   0.08 \\
  267 & 1048+0125 & 11560  &     9.62  & 1800.0  & 2750.0  & 18.0  & 12.819   &   0.03  &   11.787  &   0.07  & -23.120 & -24.152 &  1.032  &   0.08 \\
  281 & 1106+0032 &  7639  &     9.69  & 1800.0  & 2750.0  & 10.8  & 12.674   &   0.02  &   11.743  &   0.06  & -22.365 & -23.296 &  0.931  &   0.06 \\
  291 & 1116+0053 &  8059  &     9.56  & 1117.0  & 2750.0  & 18.0  & 13.128   &   0.02  &   12.211  &   0.07  & -22.028 & -22.945 &  0.917  &   0.07 \\
  294 & 1121+0058 &  7784  &     9.68  & 1800.0  & 2750.0  & 24.0  & 12.791   &   0.02  &   11.940  &   0.06  & -22.289 & -23.140 &  0.851  &   0.06 \\
  296 & 1125+0255 &  6842  &     9.81  & 1800.0  & 2000.0  & 25.0  & 12.155   &   0.04  &   11.253  &   0.06  & -22.645 & -23.547 &  0.902  &   0.07 \\
  298 & 1129+0013 & 11899  &     9.99  & 2400.0  & 3000.0  & 14.0  & 13.975   &   0.02  &   13.029  &   0.06  & -22.027 & -22.973 &  0.946  &   0.06 \\
  299 & 1130-0027 &  6688  &     9.50  & 3550.0  & 4900.0  & 14.0  & 14.053   &   0.02  &   13.134  &   0.05  & -20.698 & -21.617 &  0.919  &   0.05 \\
  310 & 1156+0142 & 14131  &     9.77  & 2250.0  & 2850.0  & 19.2  & 13.586   &   0.03  &   12.623  &   0.05  & -22.789 & -23.752 &  0.963  &   0.06 \\
  316 & 1208+0120 & 14100  &     9.72  & 1800.0  & 2750.0  & 18.0  & 14.077   &   0.05  &   13.172  &   0.06  & -22.293 & -23.198 &  0.905  &   0.08 \\
  324 & 1211+0226 & 22259  &    10.19  & 2250.0  & 2750.0  & 17.0  & 14.187   &   0.03  &   13.515  &   0.05  & -23.175 & -23.847 &  0.672  &   0.06 \\
  326 & 1213+0221 & 23252  &    10.09  & 2880.0  & 4500.0  &  5.0  & 16.700   &   0.04  &   15.762  &   0.04  & -20.757 & -21.695 &  0.938  &   0.06 \\
  327 & 1213+0127 & 14976  &     9.76  & 1200.0  & 2000.0  & 12.0  & 13.039   &   0.04  &   12.112  &   0.05  & -23.462 & -24.389 &  0.927  &   0.06 \\
  345 & 1226+0105 & 23655  &    10.26  & 2250.0  & 2750.0  & 17.0  & 13.526   &   0.03  &   12.788  &   0.05  & -23.968 & -24.706 &  0.738  &   0.06 \\
  346 & 1226+0119 &  6958  &     9.62  & 1200.0  & 1500.0  & 15.0  & 12.701   &   0.04  &   12.012  &   0.04  & -22.136 & -22.825 &  0.689  &   0.06 \\
  370 & 1300+0144 & 12264  &     9.98  & 1800.0  & 2500.0  & 25.0  & 13.171   &   0.04  &   12.089  &   0.06  & -22.896 & -23.978 &  1.082  &   0.07 \\
  377 & 1310-0019 & 12040  &     9.84  & 3600.0  & 7500.0  & 17.0  & 14.502   &   0.02  &   13.905  &   0.05  & -21.525 & -22.122 &  0.597  &   0.05 \\
  378 & 1315+0029 &  9497  &     9.94  & 3150.0  & 4000.0  & 16.0  & 14.259   &   0.03  &   13.498  &   0.05  & -21.253 & -22.014 &  0.761  &   0.06 \\
  380 & 1321+0137 & 16954  &     9.64  & 2250.0  & 3750.0  & 17.0  & 13.204   &   0.03  &   12.583  &   0.05  & -23.567 & -24.188 &  0.621  &   0.06 \\
  407 & 1401+0108 & 12901  &     9.66  & 2700.0  & 2750.0  & 17.0  & 13.697   &   0.04  &   12.965  &   0.05  & -22.480 & -23.212 &  0.732  &   0.06 \\
  410 & 1405+0006 &  7518  &     9.73  & 1800.0  & 2750.0  & 17.0  & 13.608   &   0.04  &   13.011  &   0.04  & -21.397 & -21.994 &  0.597  &   0.06 \\
  446 & 1446+0231 & 10261  &     9.63  & 3000.0  & 3750.0  & 10.4  & 16.251   &   0.05  &   15.481  &   0.04  & -19.429 & -20.199 &  0.770  &   0.06 \\
  447 & 1446+0238 & 10281  &     9.73  & 1800.0  & 1500.0  & 17.0  & 13.291   &   0.04  &   12.620  &   0.07  & -22.393 & -23.064 &  0.671  &   0.08 \\
  462 & 2303-0006 &  7468  &     9.71  & 1800.0  & 2750.0  & 30.0  & 12.163   &   0.04  &   11.465  &   0.05  & -22.827 & -23.525 &  0.698  &   0.06 \\
  463 & 2304+0155 &  5244  &     9.57  & 1800.0  & 3750.0  & 25.0  & 13.034   &   0.03  &   12.412  &   0.05  & -21.189 & -21.811 &  0.622  &   0.06 \\
  470 & 2312-0011 & 15380  &     9.98  & 1800.0  & 2750.0  & 18.0  & 13.074   &   0.03  &   12.305  &   0.04  & -23.485 & -24.254 &  0.769  &   0.05 \\
  471 & 2313+0008 &  8667  &     9.76  & 1350.0  & 2750.0  & 22.0  & 12.238   &   0.02  &   11.488  &   0.04  & -23.076 & -23.826 &  0.750  &   0.04 \\
  473 & 2315-0000 &  8938  &     9.71  & 1800.0  & 2750.0  & 22.0  & 12.212   &   0.02  &   11.398  &   0.06  & -23.168 & -23.982 &  0.814  &   0.06 \\
  474 & 2317+0112 &  9046  &     9.74  & 1350.0  & 2500.0  & 22.0  & 12.333   &   0.02  &   11.556  &   0.06  & -23.074 & -23.851 &  0.777  &   0.06 \\
  485 & 2320+0107 &  8868  &     9.84  & 1800.0  & 2750.0  & 18.0  & 13.290   &   0.03  &   12.676  &   0.05  & -22.073 & -22.687 &  0.614  &   0.06 \\
  484 & 2320+0110 &  8951  &     9.91  & 1350.0  & 2500.0  & 15.0  & 12.853   &   0.03  &   12.150  &   0.06  & -22.531 & -23.234 &  0.703  &   0.07 \\
  487 & 2325-0000 & 10147  &    10.00  & 2700.0  & 2500.0  & 18.0  & 12.558   &   0.05  &   11.502  &   0.07  & -23.098 & -24.154 &  1.056  &   0.09 \\
  515 & 2349+0248 &  5324  &     9.79  & 1800.0  & 2750.0  & 22.0  & 12.631   &   0.04  &   12.003  &   0.04  & -21.624 & -22.252 &  0.628  &   0.06 \\
  516 & 2351+0045 & 11279  &     9.59  & 3150.0  & 3000.0  & 20.0  & 15.421   &   0.04  &   14.425  &   0.06  & -20.465 & -21.461 &  0.996  &   0.07  
\enddata
\tablenotetext{a}{ Number correlative to the \citet{impey96} catalogue.}
\tablenotetext{b}{ Name from the \citet{impey96} catalogue.}
\tablenotetext{c}{ Heliocentric radial velocity in km/sec from the 21-cm HI line \citep{impey96}.}
\tablenotetext{d}{ HI mass expressed as log(M$_{HI}$/M$_\odot$) \citep{impey96}.}
\tablenotetext{e}{ Total exposure time in the corresponding filter.}
\tablenotetext{f}{ Average between the major and minor axes of the elliptical aperture used \\
        by SExtractor, in arcsecs. The value corresponds to 
        3$\sigma$ above sky, in  \\ the $J$ filter. The same elliptical aperture is used 
        for the $K_s$ image.}
\tablenotetext{g}{Calibrated aperture magnitude.}
\tablenotetext{h}{Photometric error.}
\tablenotetext{i}{Absolute magnitude computed following equation \ref{abs_magnitude}, but 
considering $v_{pec} = 0$.}
\end{deluxetable}

\newpage


\begin{center}
\underline{Figure captions} 
\end{center}

\figcaption[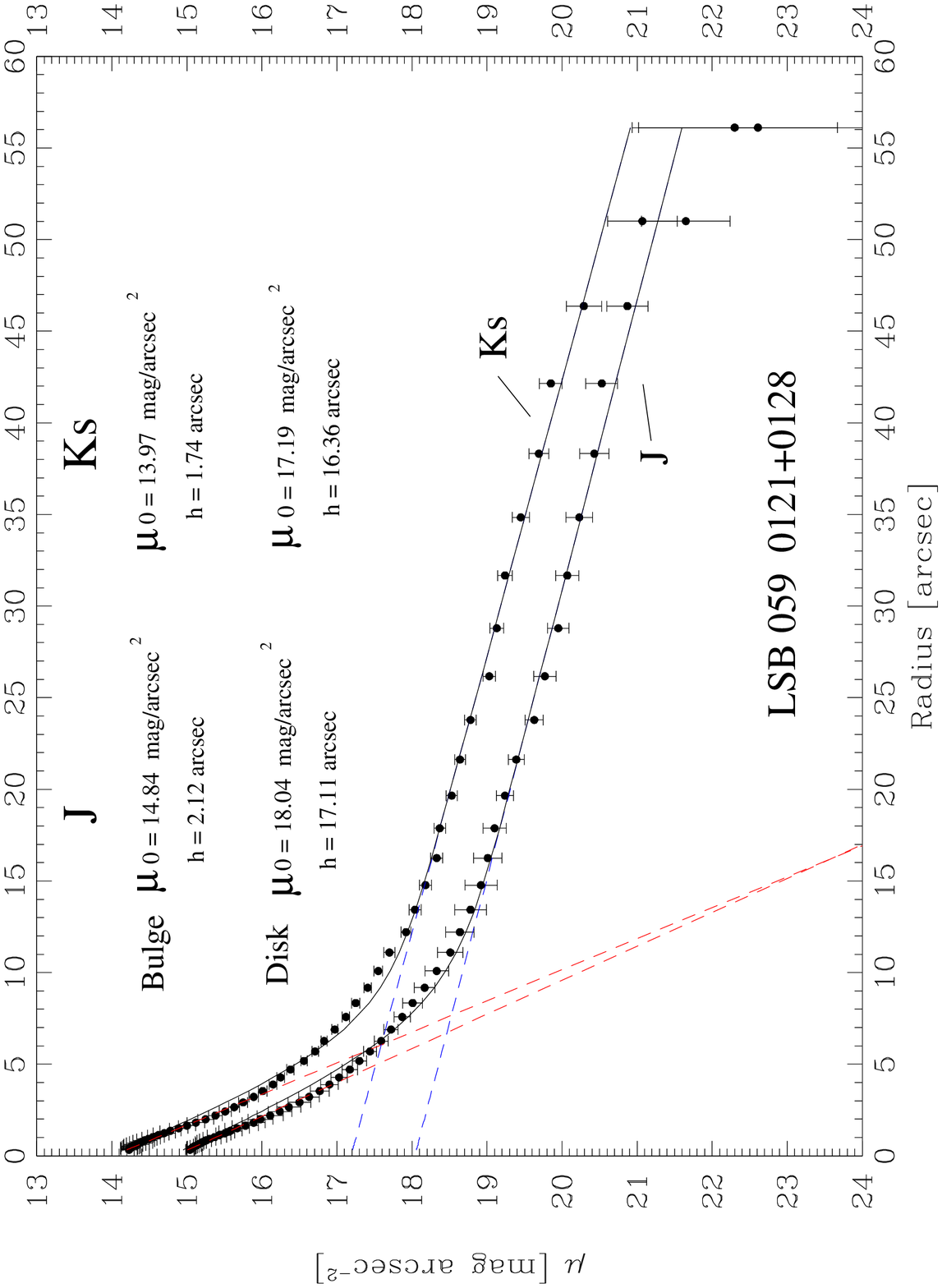]{Light profiles in the $J$ and $K_s$
bands for one LSBGs studied in the sample B. The total light profile 
is fitted by two exponential profiles, a bulge and a
disk components, using equation \ref{profile_mags}. \label{lsb059_profiles}}
\figcaption[ggalaz.fig02.eps]{Mosaic of
sample A (low HI mass sample). The label is correlative 
with column 1 of Table \ref{phot1}. Figure under request. \label{mosaics1}}
\figcaption[ggalaz.fig03.eps]{Mosaic of
sample B (high HI mass sample). The label is correlative with column 
1 of Table \ref{phot1}. Figure under request. \label{mosaics2}}
\figcaption[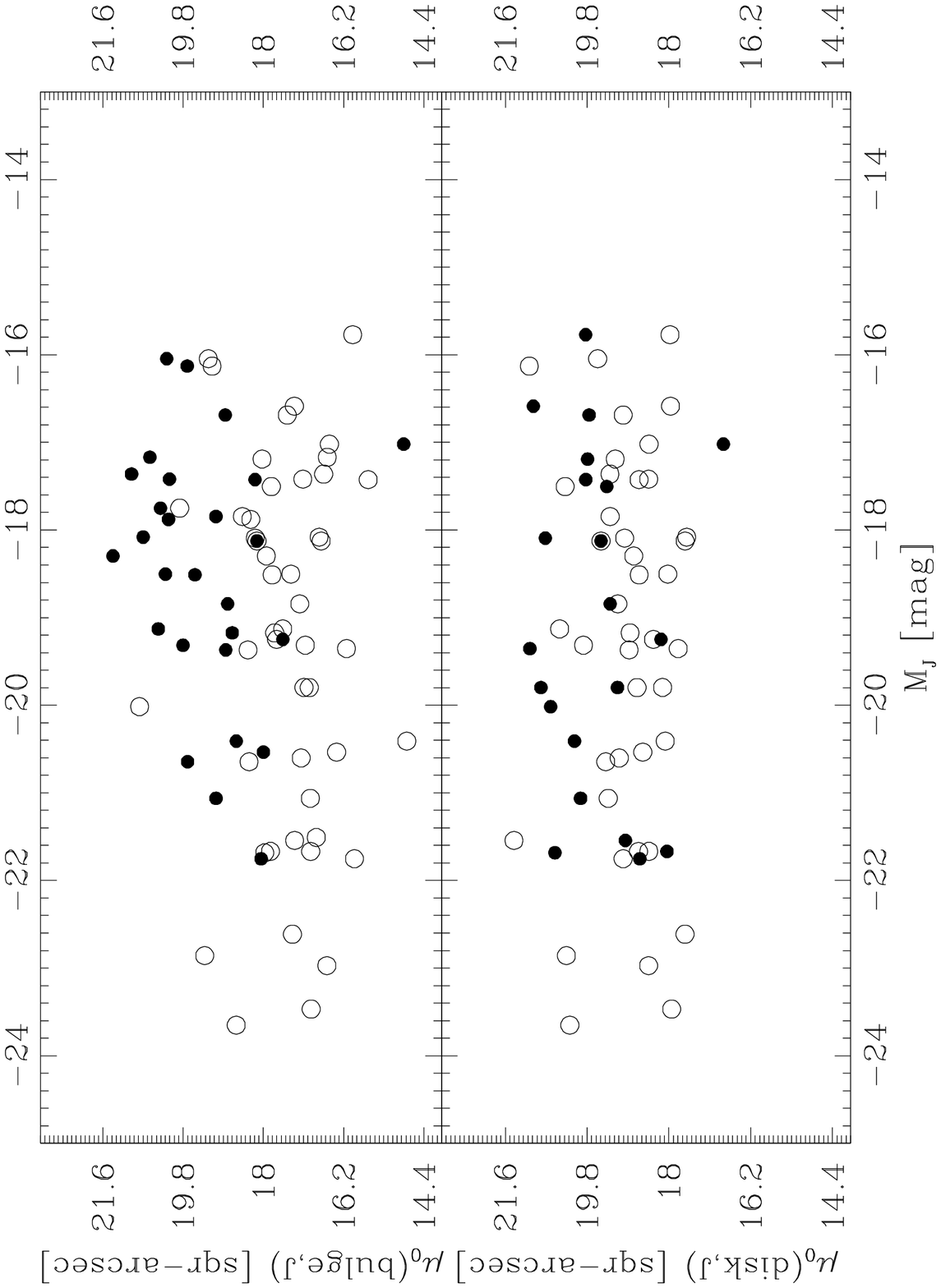]{$J$ absolute magnitude as a function of the disk 
central surface brightness. Filled circles indicate LSBGs of sample A, and
open circles LSBGs of sample B. \label{MJ_mu0}}
\figcaption[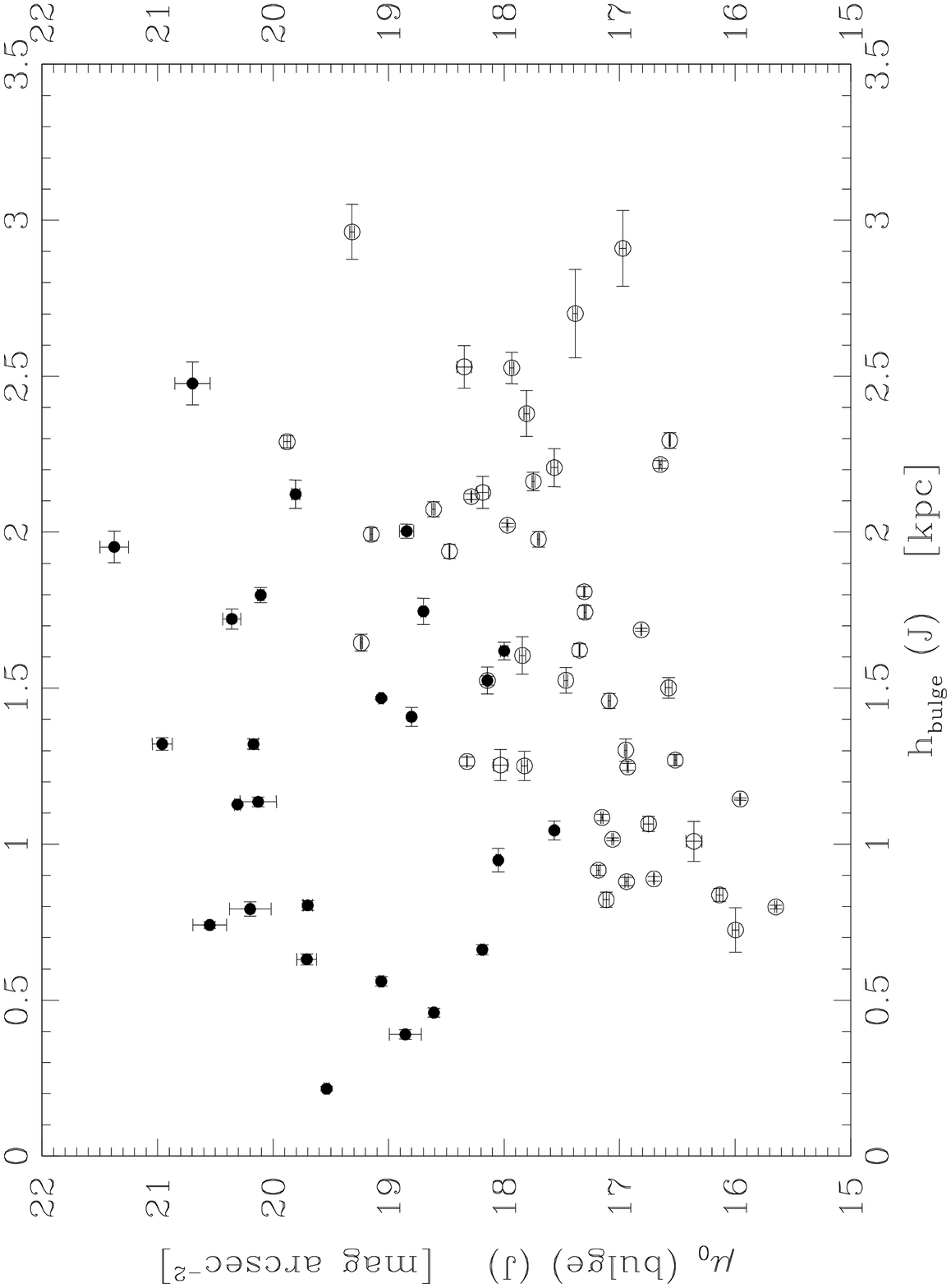,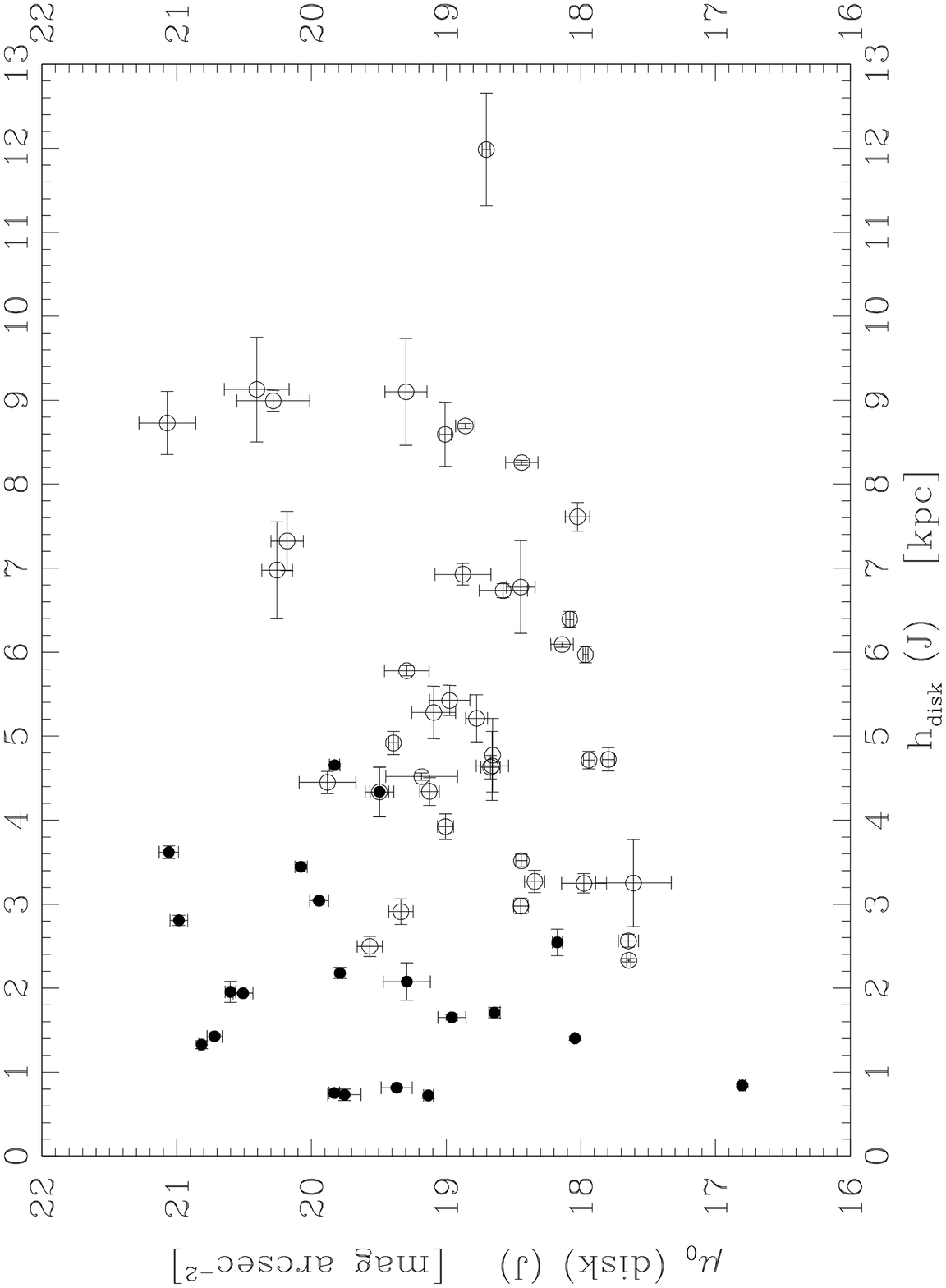]{Bulge (left) and disk
(right) central surface magnitudes as function of the respective scale
lengths in kiloparsecs. Filled circles indicate LSBGs with
log$[M_{HI}/M_{\sun}] \le 9.0$ (sample A), and open circles
log$[M_{HI}/M_{\sun}] \ge 9.5$ (sample B). \label{lsb_h_mu0_bd_J}}
\figcaption[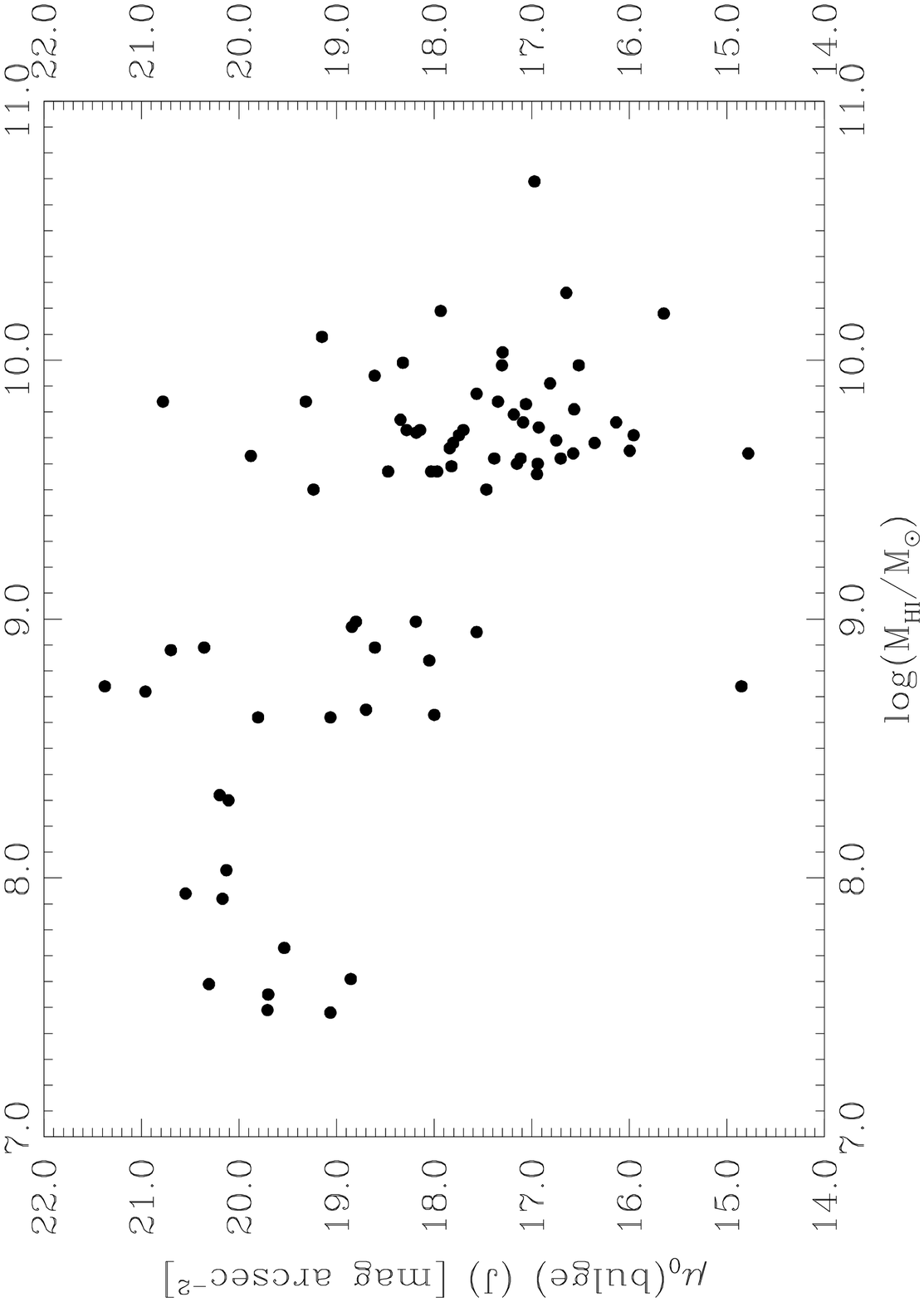,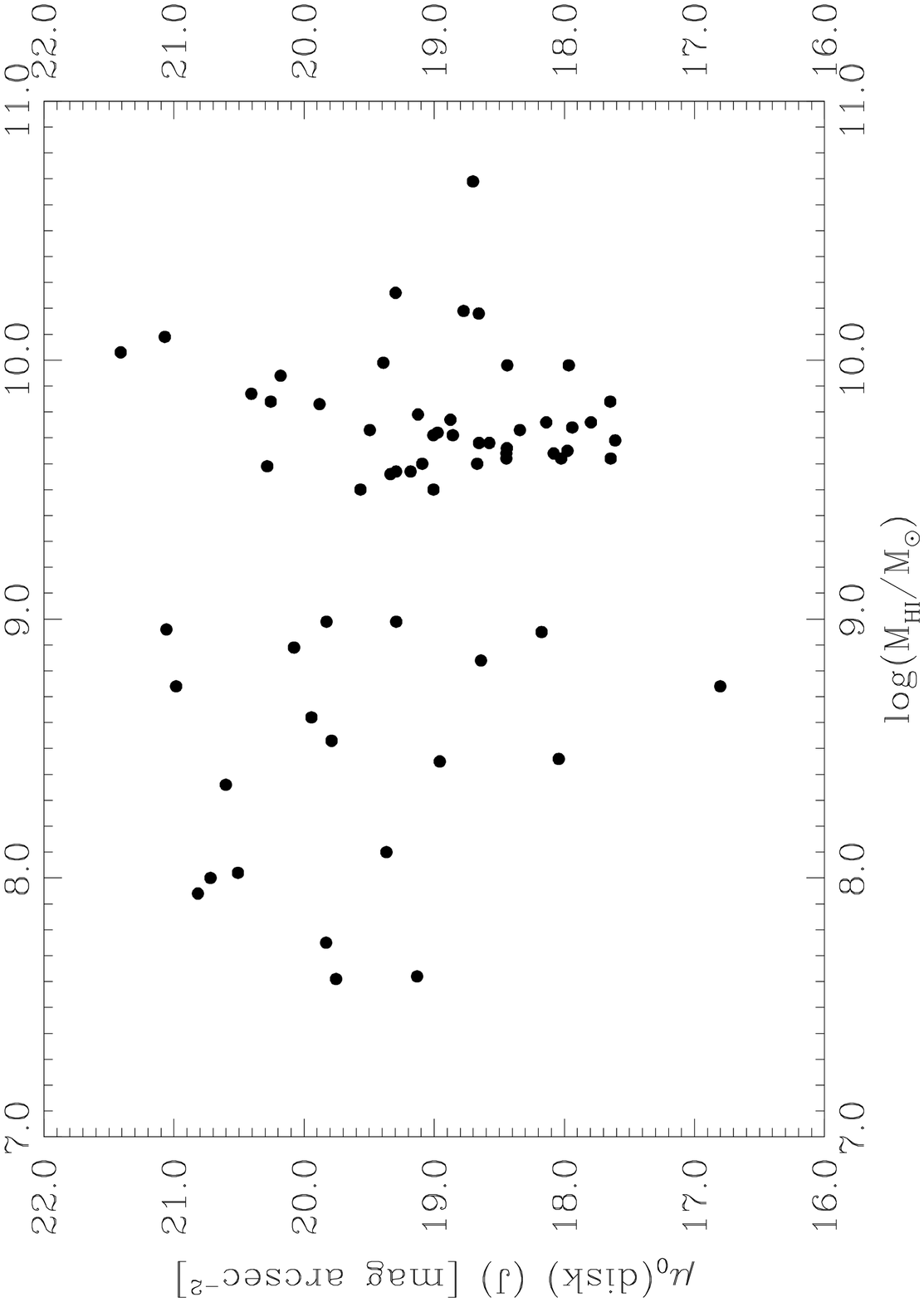]{Bulge
(left) and disk (right) central surface brightness as a function
of the HI mass. Note the large dispersion for a given HI mass.
\label{lsb_mass_mu0_J}}
\figcaption[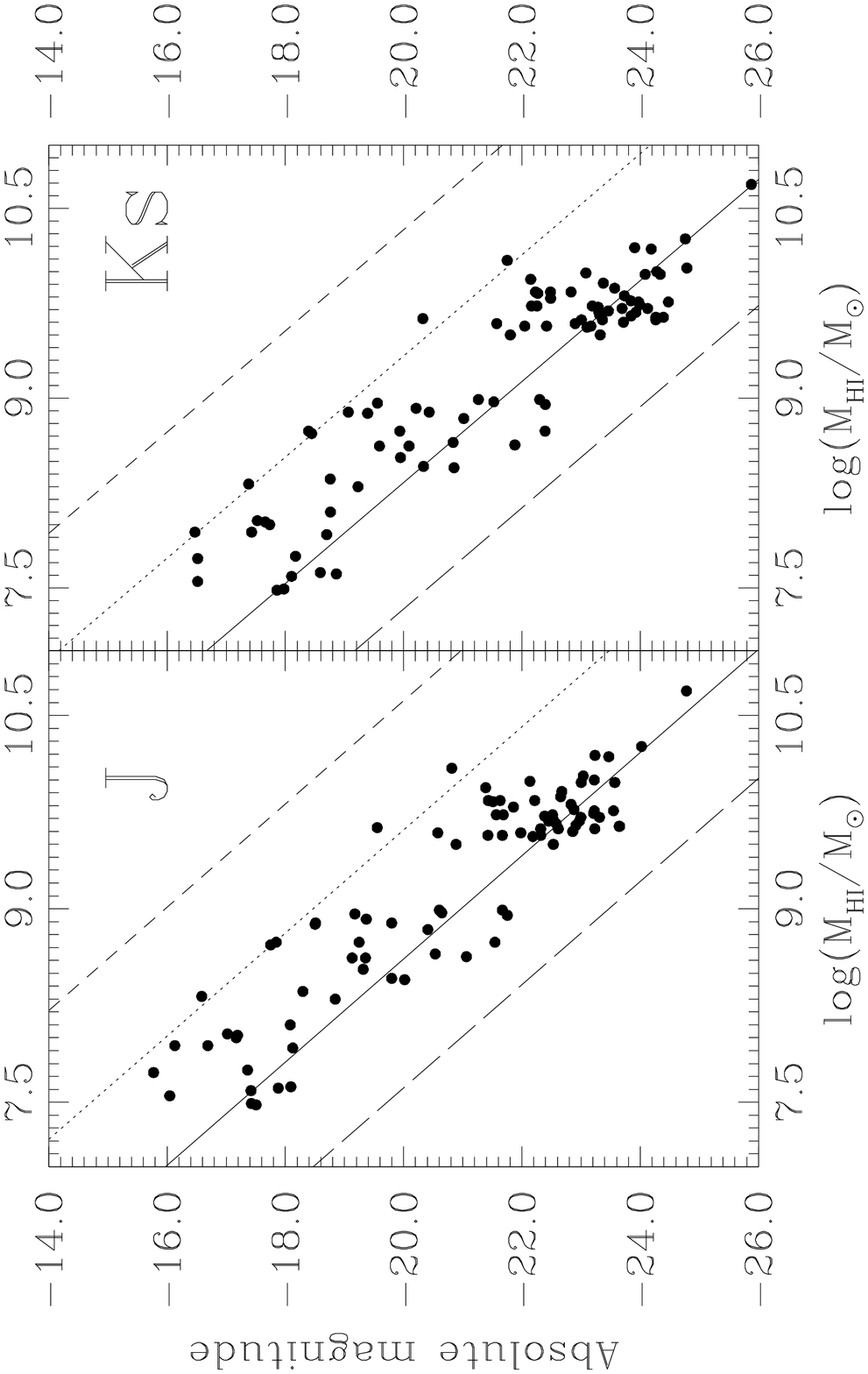]{ Absolute magnitude as a
function of HI mass. Left panel is for $J$ and right panel for $K_s$.
Diagonal lines represent constant gas fractions.  Log(M$_{HI}$/L) =
0.01 for long-dashed line, 0.1 for solid line, 1.0 for dotted line,
and 10.0 for dashed line. \label{mass_mag}}
\figcaption[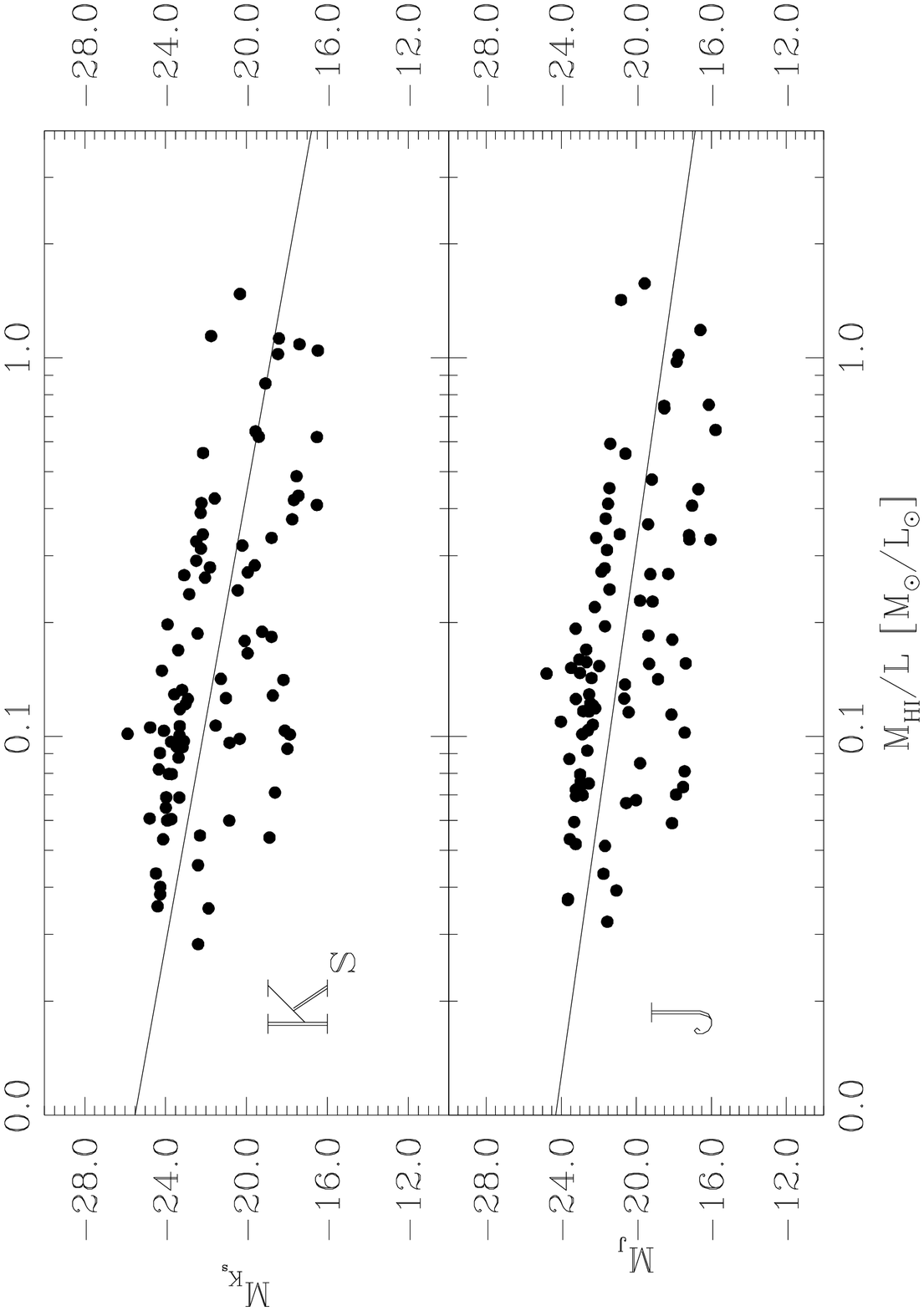]{$J$ and $K_s$ absolute magnitudes as a function
of the logarithmic gas fraction. The straight lines represent
least square fits, giving a slope of $2.86 \pm 0.46$ for $J$ and 
$3.35 \pm 0.44$ for $K_s$. \label{M_mlratio}}
\figcaption[ggalaz.fig09.eps]{$J$ images of low surface brightness galaxies
included in this paper, sorted by the gas fraction
M$_{HI}$/L$_Ks$. Galaxy 036 has a value M$_{HI}$/L$_{K_s}$ = 0.028,
and galaxy 437 M$_{HI}$/L$_{K_s}$ = 0.128. Most of these LSBGs fall
rather well in the Hubble sequence and most of them present clearly a
bulge in near-IR passbands. Figure under request. \label{sort1_ml}}
\figcaption[ggalaz.fig10.eps]{$J$ images of low surface brightness galaxies
included in this paper, sorted by the gas fraction
M$_{HI}$/L$_{K_s}$. Galaxy 225 has a value M$_{HI}$/L$_{K_s}$ = 0.129,
and galaxy 446 M$_{HI}$/L$_{K_s}$ = 1.475. Most of these LSBGs are Irr
types, but some of them are difficult to classify following the Hubble
sequence.  Most of these LSBGs do not present a bulge in the near-IR
passbands. Figure under request. \label{sort2_ml}}
\figcaption[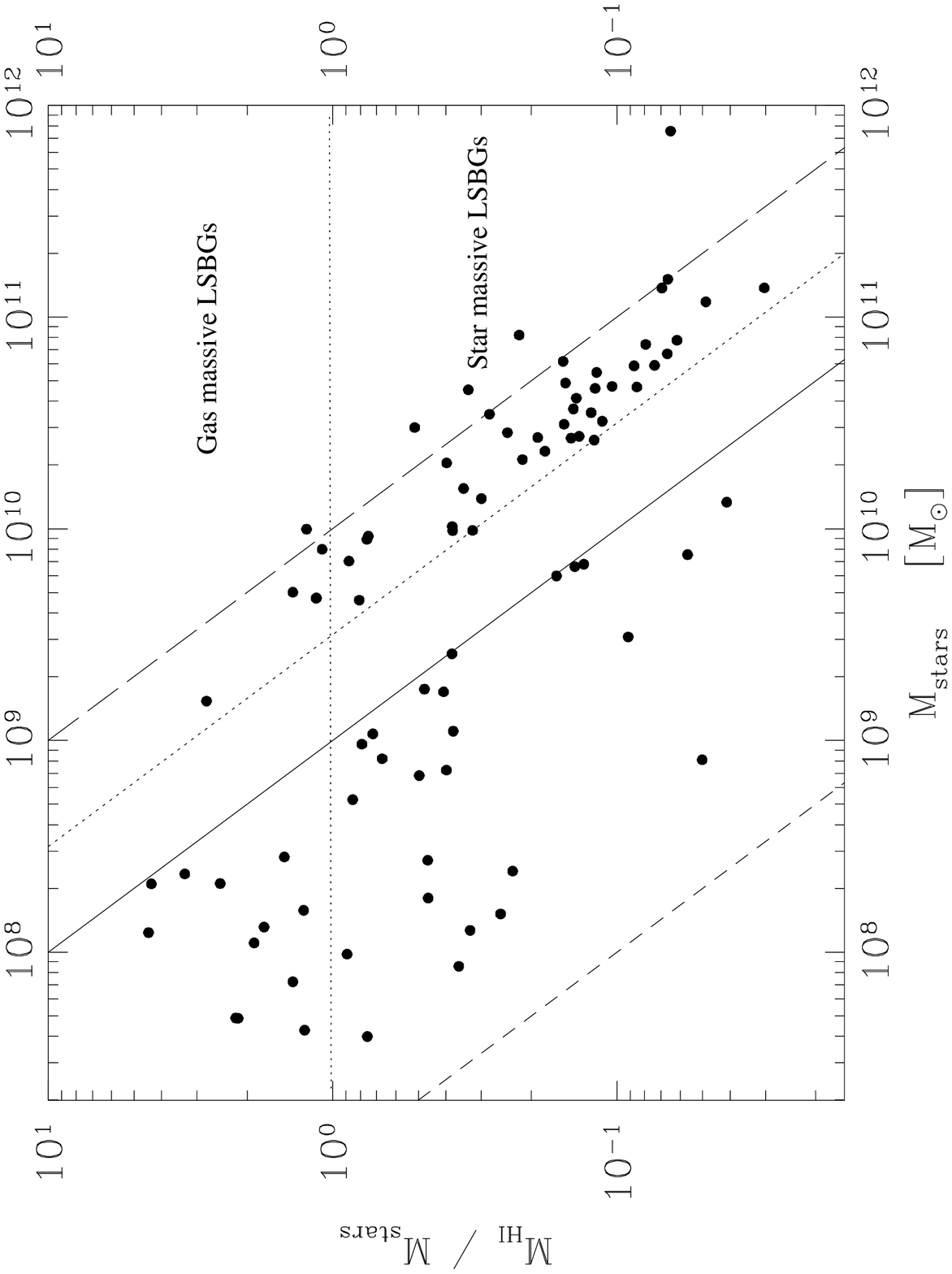]{HI-to-stellar mass ratio and color $J -
K_s$ as a function of the computed stellar mass, which has been
computed using results of \citet{bell2001}. See text for details and
discussion. Diagonal lines represent constant gas mass dashed line for
M$_{HI} = 7.0$ M$_\sun$ ( close to our lowest HI mass), solid line for
9.0 M$_\sun$ (upper limit of our sample A), dotted line for 9.5
M$_\sun$ (lower limit for our sample B), and long-dashed line for 10.0
M$_\sun$ (close to our largest HI mass). \label{Mstars_relations}}
\figcaption[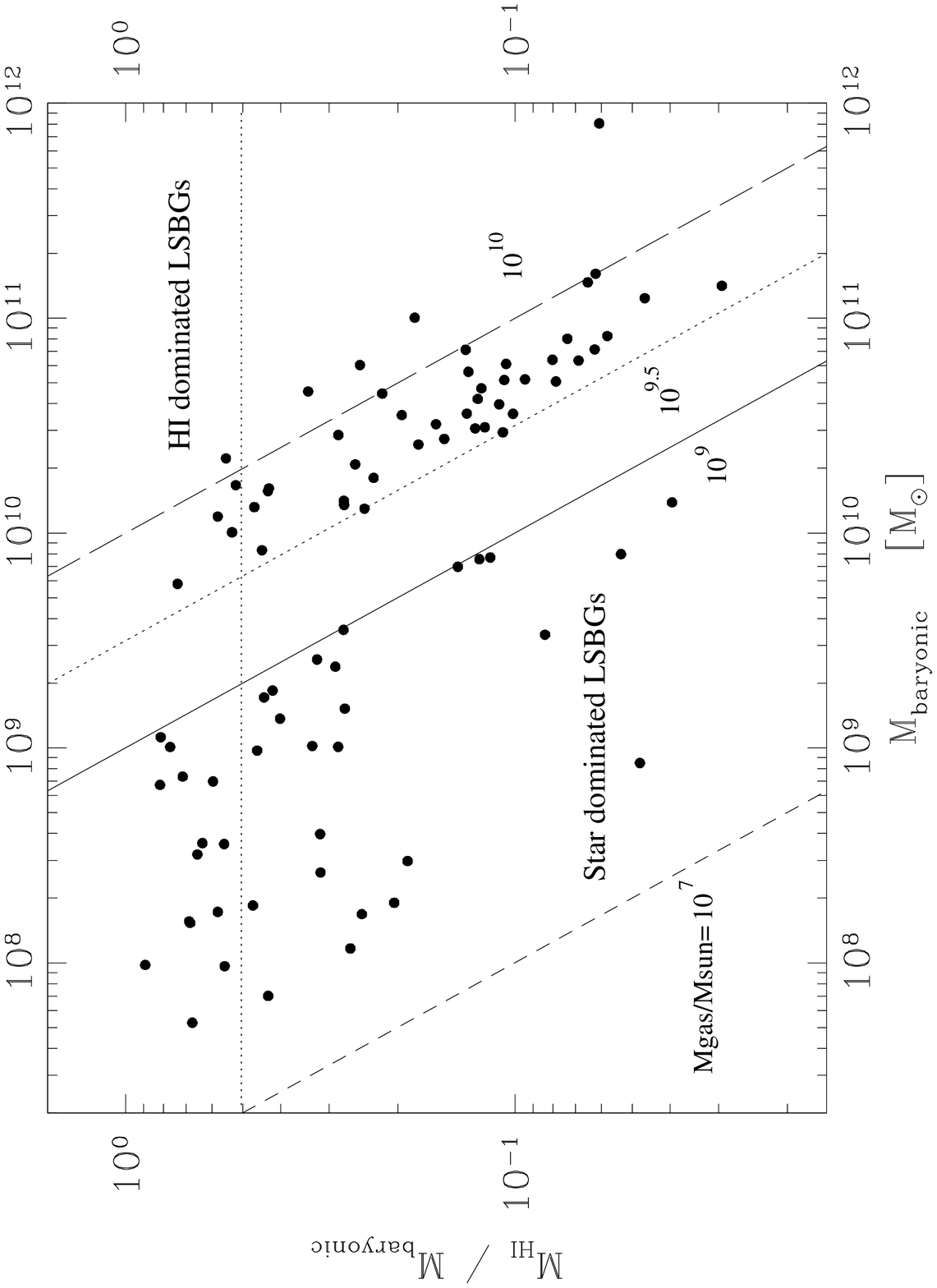]{The ratio M$_{HI}$/M$_{baryonic}$
as a function of the baryonic mass (HI mass + stellar mass). 
Diagonal lines represent constant HI content, as indicated in Figure 
\ref{Mstars_relations}. Dotted boxes indicate the locus of gas and star
dominated LSBGs. \label{Mbaryonic_relations}}
\figcaption[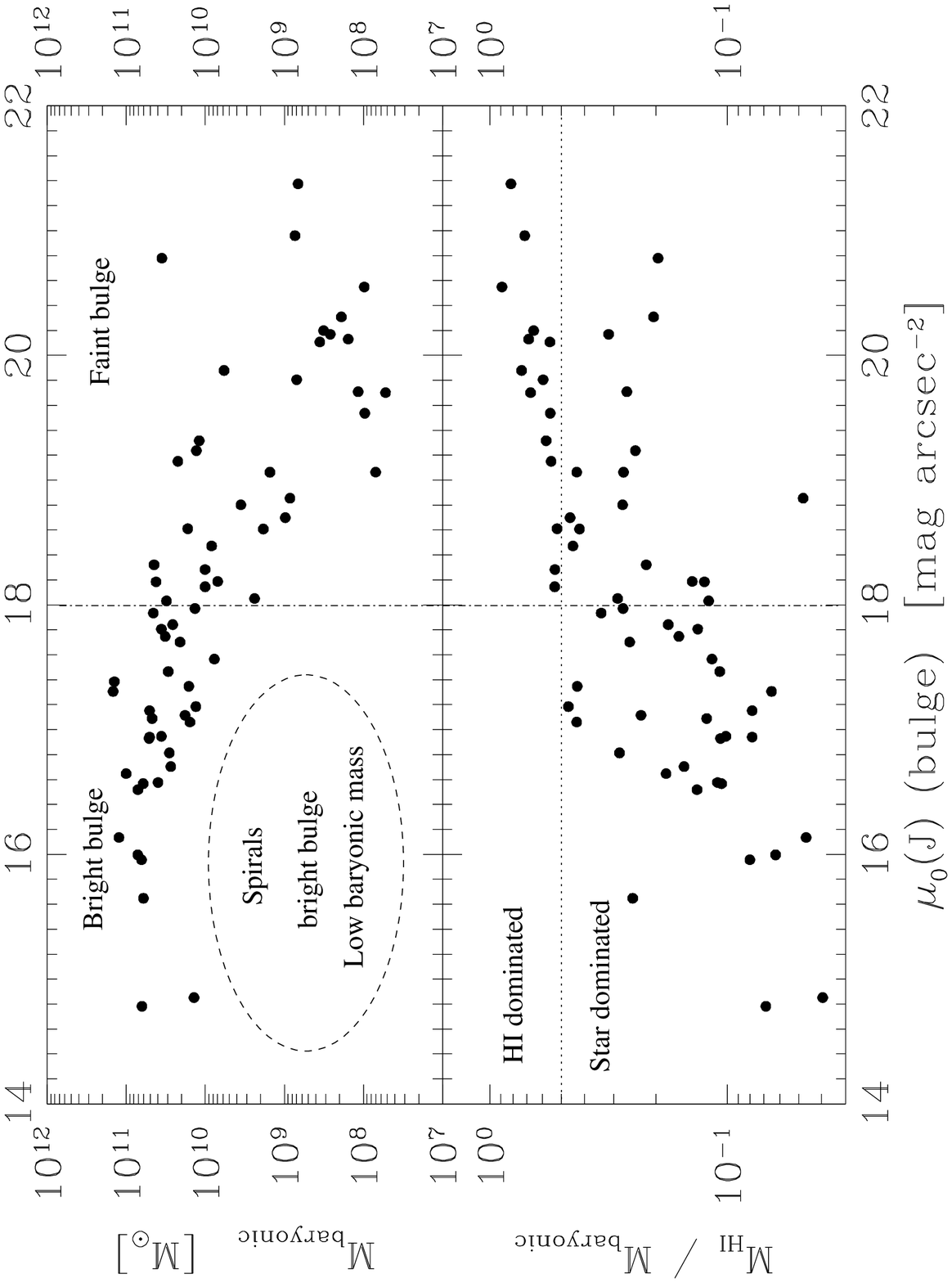]{Gas-to-stars mass fraction as a function of
the bulge surface brightness. The labels indicate the locus where gas and star
dominates and the their bulge brightness. The cutoff at $\mu_0(J,bulge) 
= 18.0$ mag arcsec$^{-2}$ represent the surface magnitude limit where 
galaxies become to be either gas or star dominated. 
See text for discussion. \label{mu0_relations}} 
\figcaption[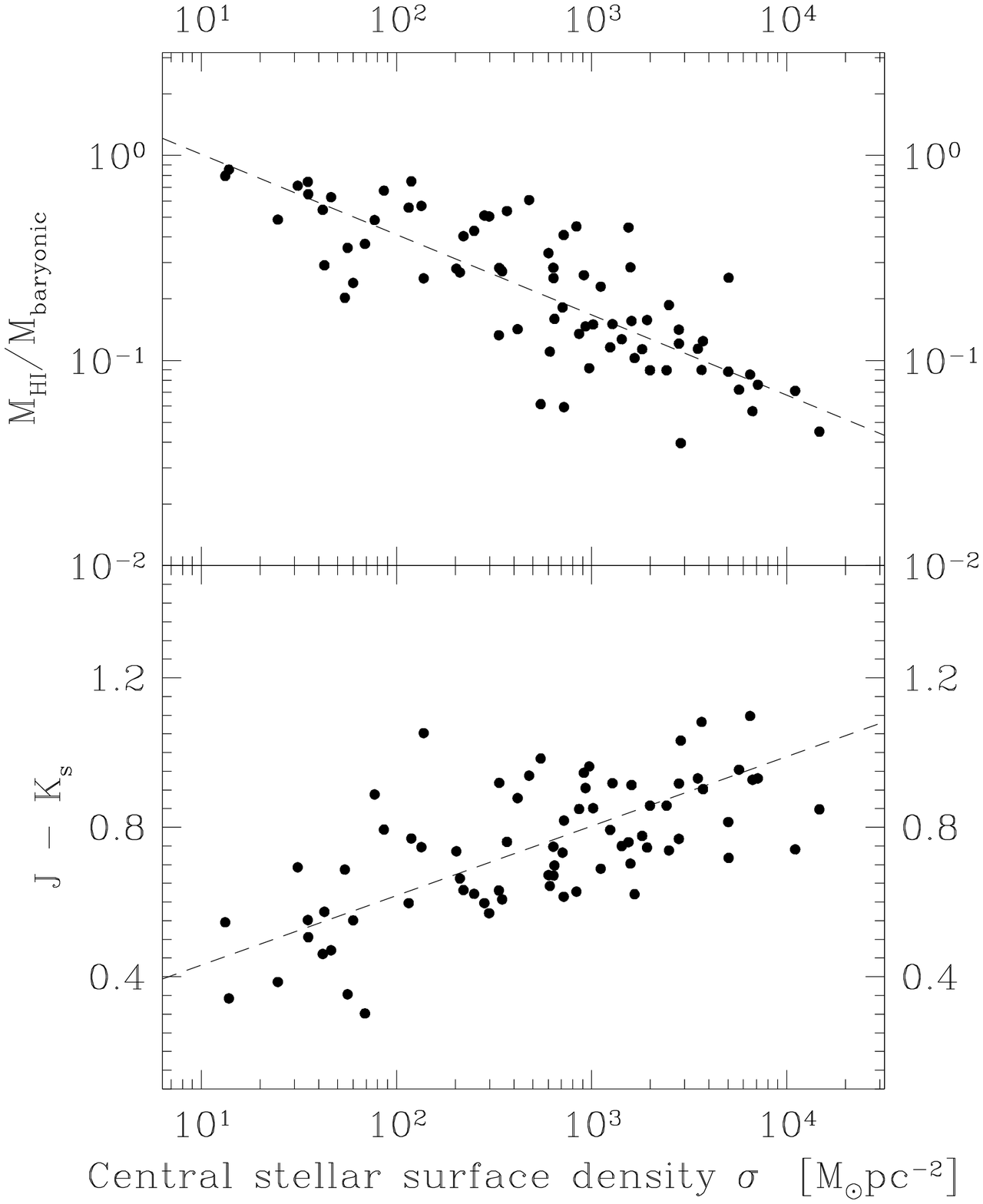]{The $J - K_s$ color and gas to baryonic
mass ratio as a function of the bulge stellar surface density
(measured in solar masses per sqr-parsec). The dashed lines represent linear
fittings: $log(M_{H_I}/M_{baryonic}) = (0.40 \pm 0.03) - (0.39 \pm 0.06) \Sigma$
and $J - K_s = (0.25 \pm 0.02) + (0.19 \pm 0.04) \Sigma$.
The correlation coefficient is 0.22 and 0.15 in the upper and lower 
panel, respectively. \label{stellar_density}} 
\figcaption[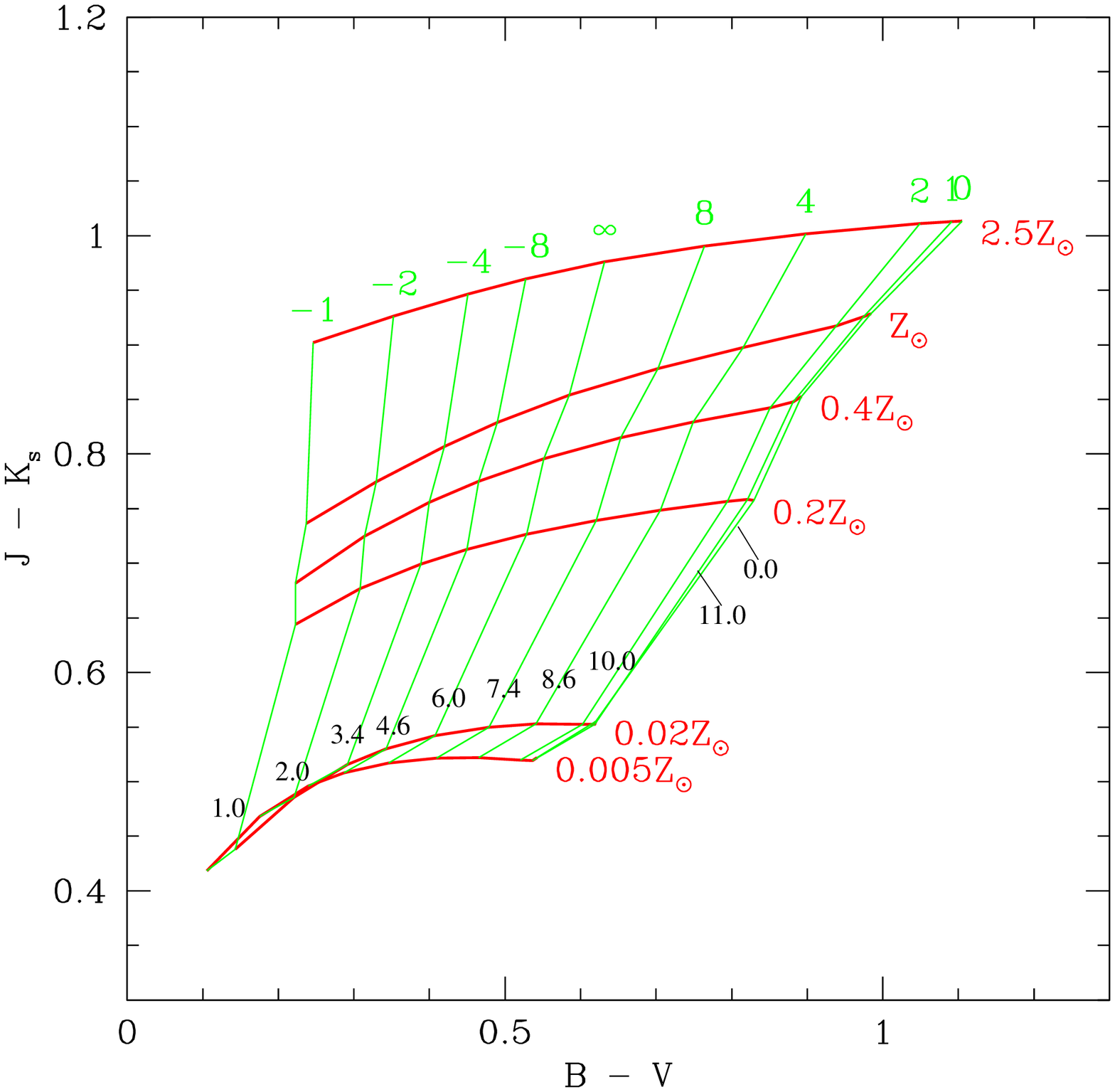]{Color grid for the $J - K_s$ index as a
function of the $B - V$ index, for different stellar formation rates
and metallicities.  Metallicities range from 0.005Z$_\sun$ to
2.5Z$_\sun$. Top labels denote different exponential star-formation
rates, where $\infty$ denotes a constant star-formation rate. Bottom
labels denote mean ages in Gyr. The star-formation started 12 Gyr ago
for all bursts.
\label{gridplot}}
\figcaption[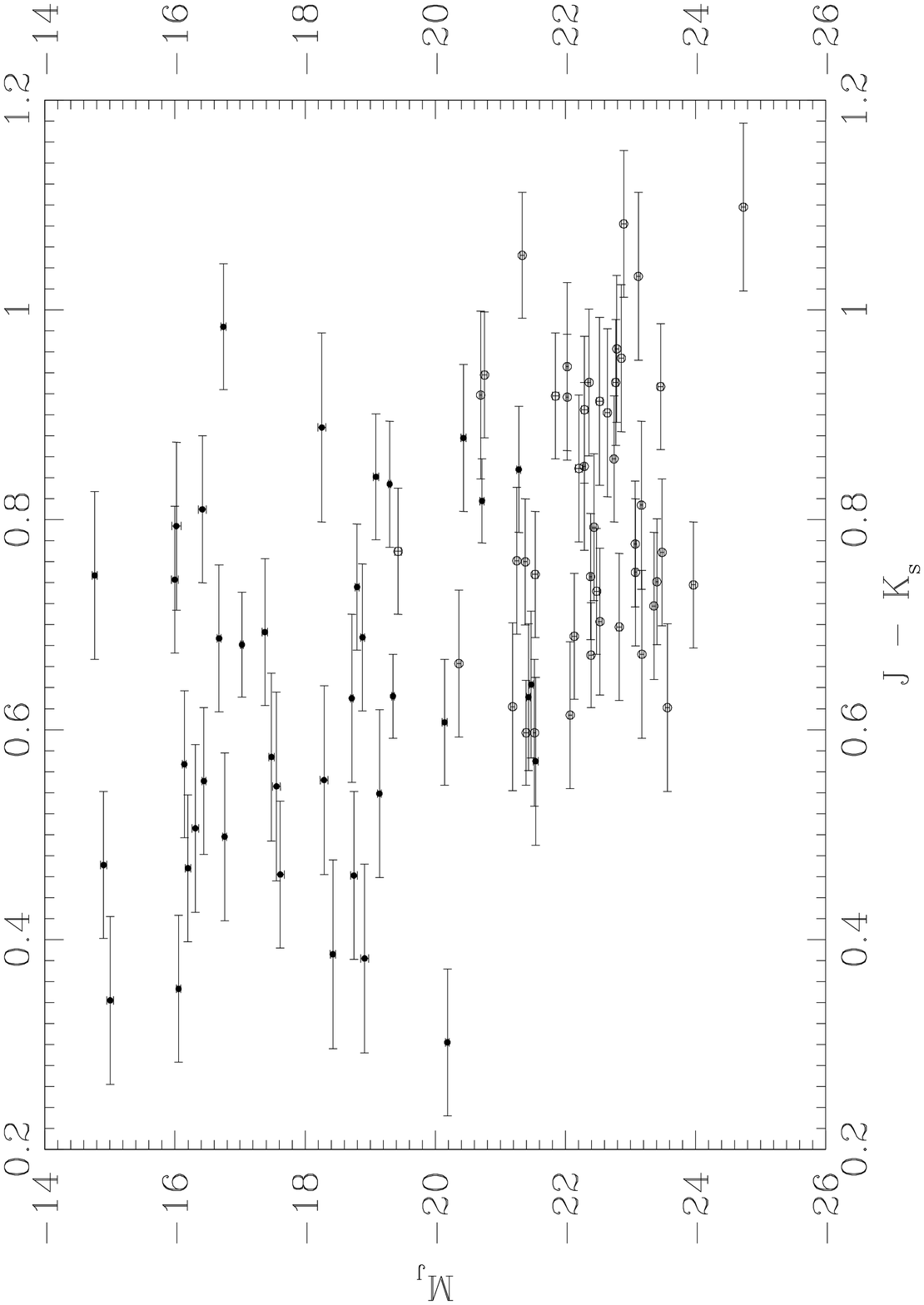,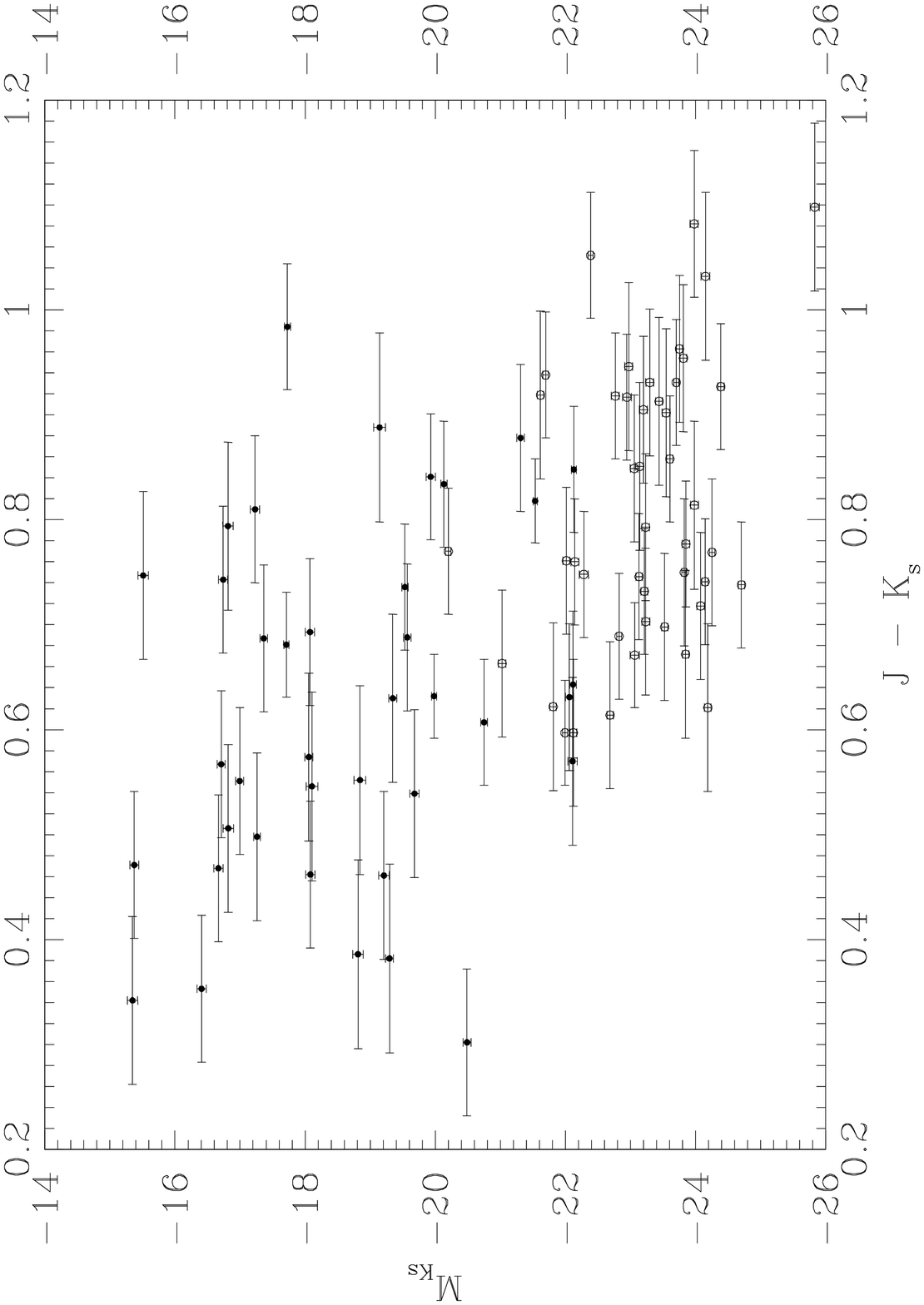]{Near-IR $J$ (left) and
$K_s$ (right) color-magnitude diagrams for LSBGs presented in this
paper.  Filled circles indicate galaxies with log$[M_{HI}/M_{\sun}]
\le 9.0$ (sample A), and open circles galaxies with
log$[M_{HI}/M_{\sun}] \ge 9.5$ (sample B). \label{abs_mag}}
\figcaption[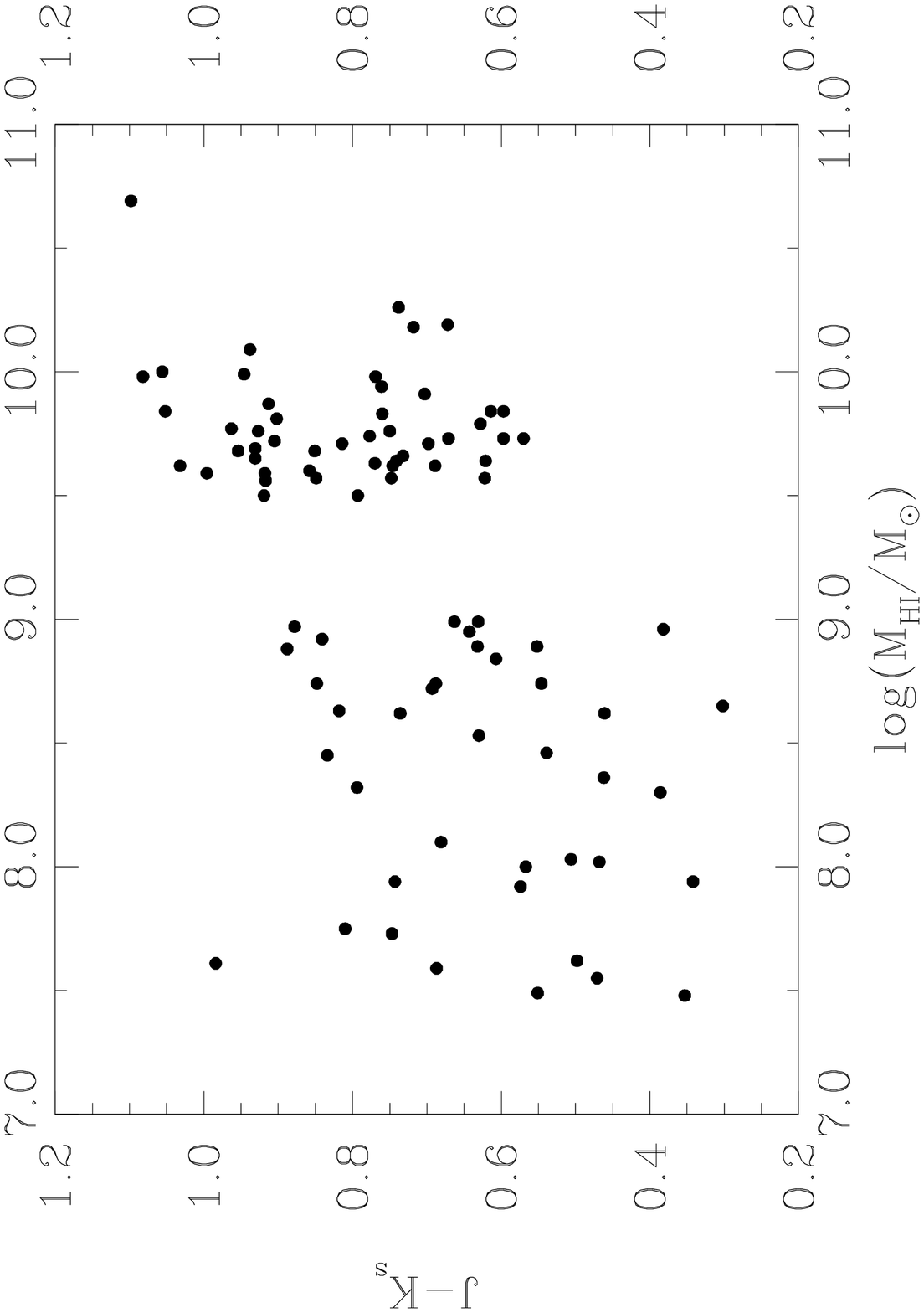]{$J-K_s$ colors as a function of the HI mass.
There is a clear positive trend in the sense of larger 
M$_{HI}$ having redder $J - K_s$. This trend is {\em not observed} in optical 
colors, as discussed in the text. \label{mass_JK}}
\figcaption[ggalaz_fig18.ps]{Computed stellar and baryonic masses as a
function of the color $J - K_s$. Straight lines represent linear
fittings. Both slopes are very similar, being $0.23 \pm 0.03$ for the
stellar masses (solid line), and $0.27 \pm 0.03$ for baryonic masses
(dashed line). \label{color_mass}}
\figcaption[ggalaz_fig19.ps]{Gas-to-baryonic mass fraction as a function 
of the $J - K_s$ color and $K_s$ absolute magnitude. Note that color become 
redder for smaller values of M$_{HI}$/M$_{baryonic}$. Dotted lines enclose
the locus of star and HI dominated LSBGs. The scatter could reflect 
a star formation process using fresh HI during the evolution to redder
colors, and so larger metallicity. See text for discussion. 
\label{ratio_lum_color}}
\figcaption[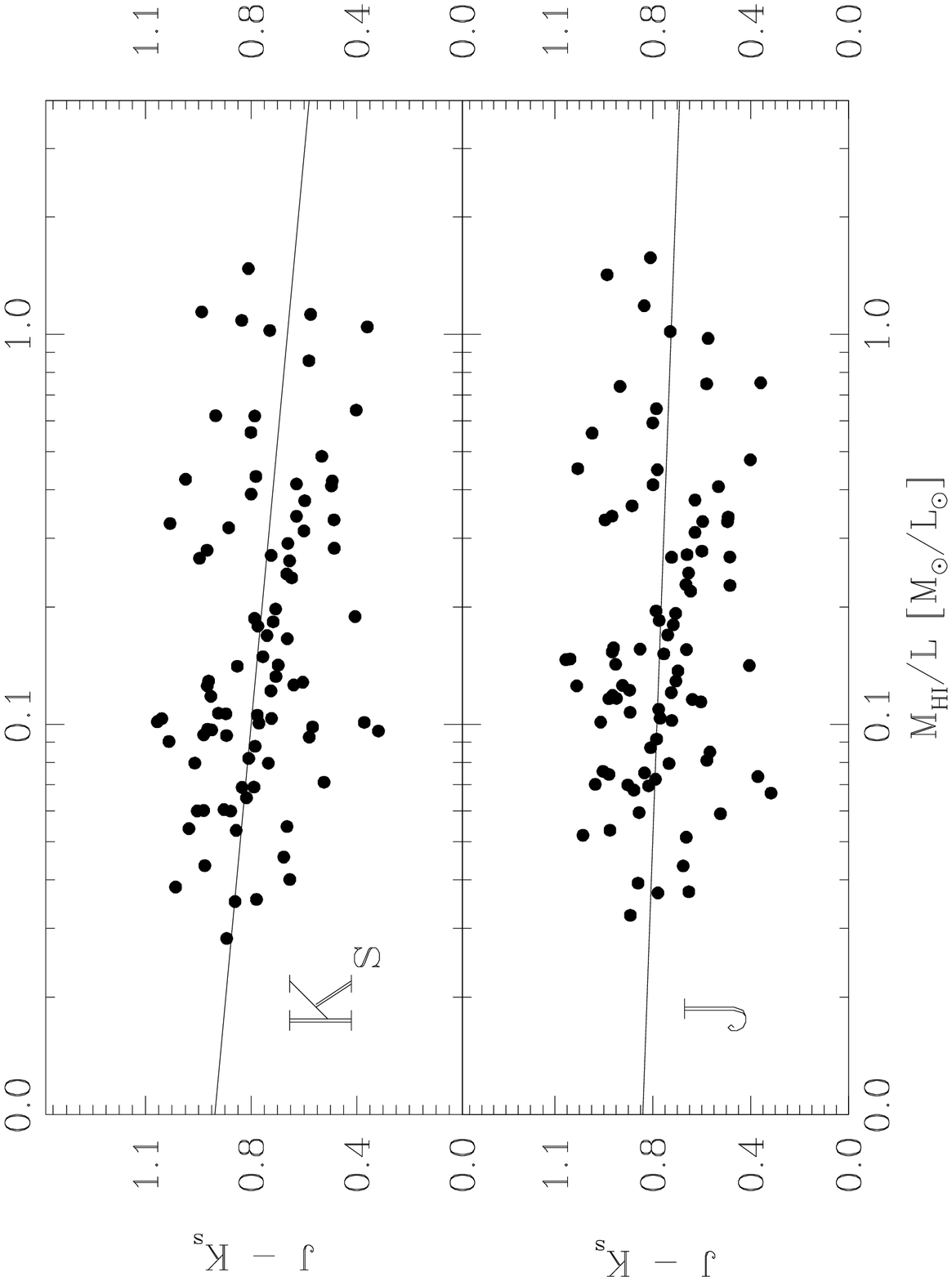]{$J - K_s$ color as a function of the 
gas fraction computed in the $J$ and $K_s$ bands.
\label{jk_mass_lum_ratio}}


\begin{figure}
\plotone{ggalaz.fig01.ps}
\figurenum{1}
\figcaption{}
\end{figure}
%
%
%
\begin{figure}
\plotone{ggalaz.fig04.ps}
\figurenum{4}
\figcaption{}
\end{figure}
\begin{figure}
\plottwo{ggalaz.fig05a.ps}{ggalaz.fig05b.ps}
\figurenum{5}
\figcaption{}
\end{figure}
\begin{figure}
\plottwo{ggalaz.fig06a.ps}{ggalaz.fig06b.ps}
\figurenum{6}
\figcaption{}
\end{figure}
\begin{figure}
\plotone{ggalaz.fig07.ps}
\figurenum{7}
\figcaption{}
\end{figure}
\begin{figure}
\plotone{ggalaz.fig08.ps}
\figurenum{8}
\figcaption{}
\end{figure}
%
%
%
\begin{figure}
\plotone{ggalaz.fig11.ps}
\figurenum{11}
\figcaption{}
\end{figure}
\begin{figure}
\plotone{ggalaz.fig12.ps}
\figurenum{12}
\figcaption{}
\end{figure}
\begin{figure}
\plotone{ggalaz.fig13.ps}
\figurenum{13}
\figcaption{}
\end{figure}
\begin{figure}
\plotone{ggalaz.fig14.ps}
\figurenum{14}
\figcaption{}
\end{figure}
\newpage
\begin{figure}
\plotone{ggalaz.fig15.ps}
\figurenum{15}
\figcaption{}
\end{figure}
\begin{figure}
\plottwo{ggalaz.fig16a.ps}{ggalaz.fig16b.ps}
\figurenum{16}
\figcaption{}
\end{figure}
\begin{figure}
\plotone{ggalaz.fig17.ps}
\figurenum{17}
\figcaption{}
\end{figure}
\begin{figure}
\plotone{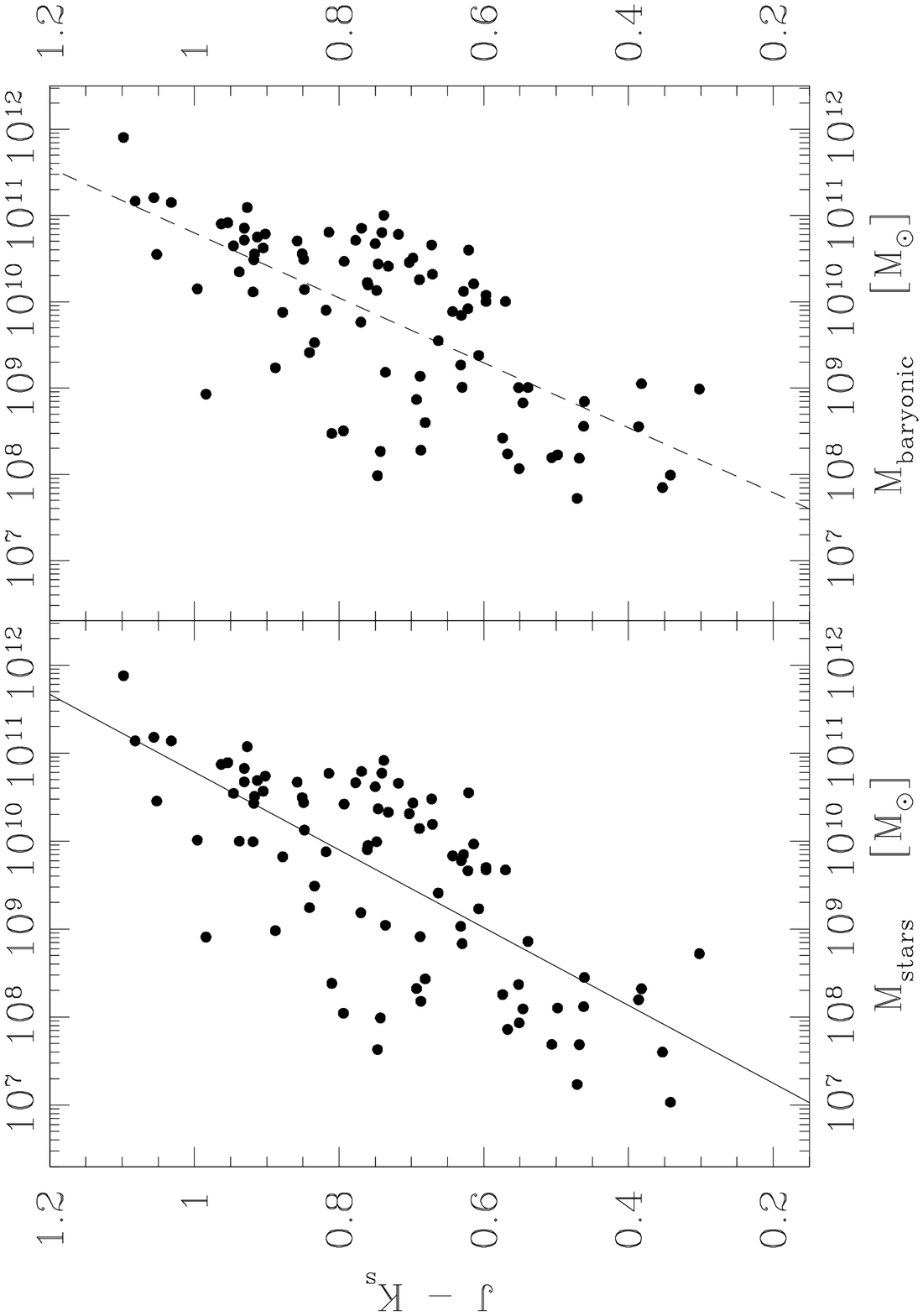}
\figurenum{18}
\figcaption{}
\end{figure}
\begin{figure}
\plotone{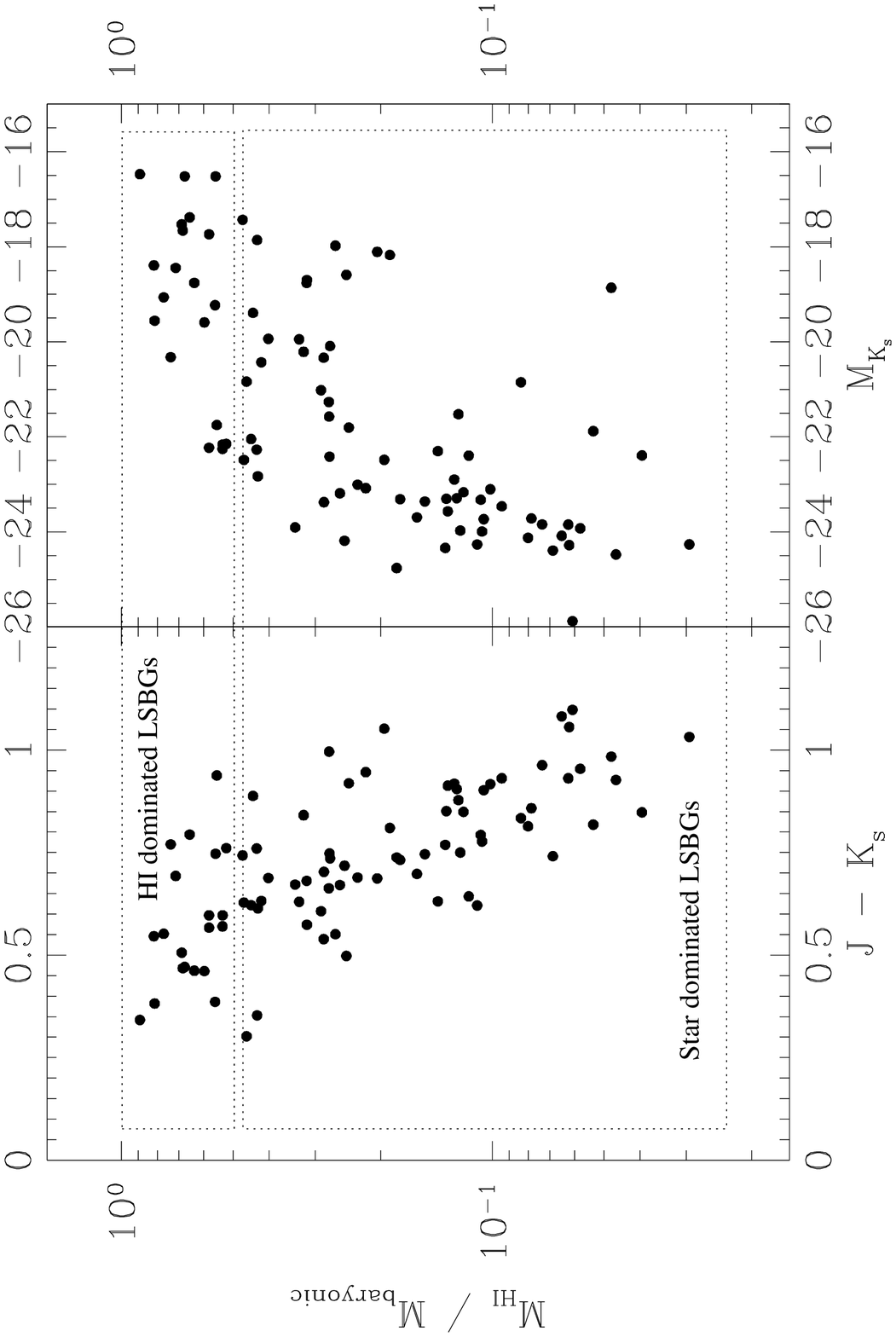}
\figurenum{19}
\figcaption{}
\end{figure}
\begin{figure}
\plotone{ggalaz.fig20.ps}
\figurenum{20}
\figcaption{}
\end{figure}

\end{document}